\documentclass{icmart}

\contact[matbaowz@nus.edu.sg; URL: http://www.math.nus.edu.sg/\~{}bao/]
{Department of Mathematics, National University of Singapore,
Singapore, 119076, Singapore}

\newtheorem{theorem}{Theorem}[section]

\newtheorem{lemma}[theorem]{Lemma}

\theoremstyle{definition}

\newtheorem{remark}[theorem]{Remark}

\newcommand{\bx}{{\bf x}}
\newcommand{\btd}{\nabla}
\newcommand{\vep}{\varepsilon}
\newcommand{\dt}{\delta}
\newcommand{\tbx}{\widetilde{\bf x} }

\title[Models and Methods for Bose-Einstein Condensation]
{Mathematical Models and Numerical Methods for Bose-Einstein Condensation}

\author[Weizhu Bao]
{Weizhu Bao\thanks{The author is grateful to Beijing Computational Science Research
Center for its hospitality during the writing of this paper.}}

\begin{document}

\begin{abstract}
The achievement of Bose-Einstein condensation (BEC) in ultracold vapors of alkali atoms
has given enormous impulse to the theoretical and experimental
study of dilute atomic gases in
condensed quantum states inside magnetic traps and optical lattices.
This article offers a short survey on mathematical models and theories as well as
numerical methods for BEC based on the mean field theory.
We start with the Gross-Pitaevskii equation (GPE) in three dimensions (3D) for
modeling one-component BEC of the weakly interacting bosons,
scale it to obtain a three-parameter model and show how to reduce it to two dimensions (2D) and
one dimension (1D) GPEs in certain limiting regimes. Mathematical theories and numerical
methods for ground states and dynamics of BEC are provided.
Extensions to GPE with an angular momentum rotation term for a rotating BEC,
to GPE with long-range anisotropic dipole-dipole interaction for a dipolar BEC
and to coupled GPEs for spin-orbit coupled BECs are discussed.
Finally, some conclusions are drawn and future research
perspectives are discussed.
\end{abstract}



\maketitle

\section{Introduction}
The achievement of Bose-Einstein condensation (BEC)
of dilute gases in 1995 \cite{Anderson,Bradley,Davis} marked the beginning of
a new era in atomic, molecular and optical (AMO) physics
and quantum optics. In fact, the phenomenon known as BEC
was predicted by Einstein in 1924 \cite{Einstein1,Einstein2} based on
the ideas of Bose  \cite{Bose} concerning photons:
In a system of bosons obeying Bose statistics under the assumption that it
is in equilibrium at temperature $T$ and chemical potential $\mu$,
Einstein \cite{Einstein1,Einstein2} derived the so-called
Bose-Einstein distribution (or Bose-Einstein statistics),
in the grand canonical ensemble, for the mean occupation of the $j$th energy state as
\begin{equation}
\numberwithin{equation}{section}
n_j=\frac{1}{e^{(\varepsilon_j-\mu)/k_BT}-1}:= f(\varepsilon_j), \qquad j=0,1,\ldots\;,
\end{equation}
where $\varepsilon_j>\mu$ is the energy of the $j$th state,
$n_j$ is the number of particles in state $j$,
$k_B$ is the Boltzmann constant. The mean total number of particles is given as
$N(T,\mu)=\sum\limits_{j=0}^\infty f(\varepsilon_j)$, and the mean total energy is given as
$E(T,\mu)=\sum\limits_{j=0}^\infty \varepsilon_jf(\varepsilon_j)$.
From the above distribution, Einstein \cite{Einstein1,Einstein2} predicted
that there should be a critical temperature $T_c$ below which a finite fraction
of all the particles ``condense'' into the same one-particle state.

Einstein's original prediction was for a noninteracting gas and did not receive much
attention in a long time. After the observation of superfluidity in liquid $^4$He
below the $\lambda$ temperature ($2.17$K) in 1938, London \cite{London}
suggested that despite the strong interatomic interactions
BEC was indeed occurring in this system and was responsible for
the superfluid properties. This suggestion has stood the test
of time and is the basis for our modern understanding of the properties
of the superfluid phase. By combining laser cooling and evaporative cooling,
in 1995 BEC was realized in a system that is about as different as possible from
$^4$He, namely, dilute atomic alkali gases trapped by magnetic fields
and over the last two decades these systems have been the subject of an explosion
of research, both experimental and theoretical. Perhaps the single
aspect of BEC systems that makes them most fascinating is best illustrated
by the cover of {\sl Science} magazine of December 22, 1995,
in which the Bose condensate is declared ``{\bf molecule of the year}'' and
pictured as a platoon of soldiers marching in lock-step:
every atom in the condensate must behave in exactly the same way, and
this has the consequence, {\sl inter alia}, that effects which are so small as to be
essentially invisible at the level of single atom may be spectacularly amplified.
Most BEC experiments reach quantum degeneracy between $50$ nK and $2$ $\mu$K, at densities
between $10^{11}$ and $10^{15}$ cm$^{-3}$. The largest condensates are of
$100$ million atoms for sodium, and a billion for hydrogen; the smallest are just a
few hundred atoms. Depending on the magnetic trap, the shape of the condensate is
either approximately round, with a diameter of $10$--$15$ $\mu$m, or cigar-shaped with about
$15$ $\mu$m in diameter and $300$ $\mu$m in length. The full cooling cycle that
produce a condensate may take from a few seconds to as long as several minutes \cite{Cornell,Ketterle}.
For better understanding of the long history towards the BEC and its physical study,
we refer to the Nobel lectures \cite{Cornell,Ketterle} and several review papers
in physics \cite{Dalfovo,Leggett,Pethick,PitaevskiiStringari}.

The experimental advances in BEC \cite{Anderson,Bradley,Davis}  have spurred
great excitement in the AMO community and condense matter community as well as
computational and applied mathematics community.
Since 1995, numerous efforts have been devoted to the studies of ultracold atomic gases
and various kinds of condensates of dilute gases have been produced for
both bosonic particles and fermionic particles
\cite{Dalfovo,Fetter,Leggett}.
In this rapidly growing research area,   mathematical models and analysis as well as
numerical simulation have been playing an important role  in  understanding
the theoretical part of BEC and predicting and guiding the experiments.
The goal of this paper is to offer a short survey on mathematical models and theories as well as
numerical methods for BEC based on the Gross-Pitaevskii equation (GPE)
\cite{Baocai2013,Gross,Pethick,Pitaevskii,PitaevskiiStringari}.
The paper is organized as follows. In section 2, we present
the GPE for BEC based on the mean field approximation. Ground states and their computations
are discussed in section 3, and dynamics and its computation are presented in section 4.
Extensions to
rotating BEC, dipolar BEC and spin-orbit-coupled BEC are presented in
section 5. Finally, some  conclusions and perspectives are drawn in section
6.

\section{The Gross-Pitaevskii equation}
In this section, we will present the GPE for modeling BEC based on the mean field approximation
\cite{Baocai2013,Gross,Pethick,Pitaevskii,PitaevskiiStringari},
its nondimensionalization and dimension reduction to lower dimensions.

\subsection{Mean field approximation}
For a BEC of ultracold dilute gas with $N$ identical bosons
confined in an external trap, only binary interaction is important, then
the many-body Hamiltonian for it
can be written as \cite{LiebSeiringer,Leggett}
\begin{equation}\label{Ham}
H_{N}=\sum\limits_{j=1}^N\left(-\frac{\hbar^2}{2m}\Delta_j+V(\bx_j)\right)+
\sum\limits_{1\leq j<k\leq N}V_{\rm int}(\bx_j-\bx_k),
\end{equation}
where $\bx_j\in\Bbb R^3$ denotes the position of the $j$th particle for $j=1,\ldots,N$,
$m$ is the mass of a boson, $\hbar$ is the Planck constant, $\Delta_j=\nabla_j^2$
is the Laplace operator with
respect to $\bx_j$, $V(\bx_j)$ is the external trapping potential,
and $V_{\rm int}(\bx_j-\bx_k)$ denotes the inter-atomic two body interaction.
Denote the complex-valued wave function $\Psi_N:=\Psi_N(\bx_1,\ldots,\bx_N,t)\in
L^2(\Bbb R^{3N}\times \Bbb R)$  for the $N$ particles in the BEC,
which is symmetric with respect to any permutation
of the positions $\bx_j$ ($1\le j\le N$), then the total energy is given as
\begin{equation}\label{energy-total}
E_{\rm total}(\Psi_N)=(\Psi_N,H_N\Psi_N):=
\int_{{\Bbb R}^{3N}}\overline{\Psi}_N H_N\Psi_N \; d\bx_1\ldots d\bx_N,
\end{equation}
where $\overline{f}$, ${\rm Re}(f)$ and ${\rm Im}(f)$ denote the complex conjugate,
real part and imaginary part of $f$, respectively, and
the evolution of the system is described by
the time-dependent linear Schr\"odinger equation
\begin{equation}\label{eq:Nbdy}
i\hbar\partial_t\Psi_N(\bx_1,\ldots,\bx_N,t)=\frac{\delta E_{\rm total}(\Psi_N)}{\delta \overline{\Psi}_N}=H_N\Psi_N(\bx_1,\ldots,\bx_N,t),
\end{equation}
where $i=\sqrt{-1}$ denotes the imaginary unit and $t$ is time.

For a BEC, all  particles are in the same quantum state and we can formally
take the Hartree ansatz \cite{Baocai2013,Erdos,Gross,LiebSeiringer,LiebSeiringerPra2000,Pethick,Pitaevskii,PitaevskiiStringari}
\begin{equation}\label{eq:HF}
\Psi_N(\bx_1,\ldots,\bx_N,t)\approx \prod_{j=1}^N\psi(\bx_j,t),
\end{equation}
with the normalization for the single-particle wave function $\psi:=\psi(\bx,t)$ as
\begin{equation} \label{GPnorm}
\|\psi(\cdot,t)\|^2:=\int_{\Bbb R^3}|\psi(\bx,t)|^2\,d\bx=1,
\end{equation}
where $\bx=(x,y,z)^T\in {\Bbb R}^3$ is the Cartesian coordinate in three dimensions (3D).
Due to that the BEC gas is dilute and the temperature is below the critical temperature $T_c$, i.e.
a weakly interacting gas,
the binary interaction $V_{\rm int}$ is well approximated by the effective
contact interacting potential \cite{Pethick,Pitaevskii,PitaevskiiStringari}:
\begin{equation}\label{eq:mf-app}
V_{\rm int}(\bx_j-\bx_k)=g\,\delta(\bx_j-\bx_k),
\end{equation}
where $\delta(\cdot)$ is the Dirac distribution and the constant $g=\frac{4\pi\hbar^2 a_s}{m}$
with $a_s$  the $s$-wave scattering length of the bosons
(positive for repulsive interaction and negative for attractive interaction,
which is much smaller than the average distance between the particles).
Plugging (\ref{eq:HF}) into (\ref{energy-total}),
noticing (\ref{Ham}) and (\ref{eq:mf-app}), and keeping only the two-body interaction,
we obtain $E_{\rm total}(\Psi_N)\approx  N\,E(\psi)$ with the Gross-Pitaevskii (GP) energy
(or energy per particle) defined as \cite{Gross,LiebSeiringer,LiebSeiringerPra2000,Pethick,Pitaevskii,PitaevskiiStringari}
\begin{equation} \label{GPenergy}
E(\psi)=\int_{\Bbb R^3}\left[\frac{\hbar^2}{2m}|\nabla \psi(\bx,t)|^2+V(\bx)|\psi(\bx,t)|^2
+\frac{Ng}{2}|\psi(\bx,t)|^4\right]\,d\bx.
\end{equation}
The dynamics of the BEC will be governed by the following nonlinear Schr\"odinger equation (NLSE) with cubic nonlinearity, known as the Gross-Pitaevskii equation (GPE) \cite{Baocai2013,Erdos,Gross,LiebSeiringer,LiebSeiringerPra2000,Pethick,Pitaevskii,PitaevskiiStringari}:
\begin{equation}\label{eq:GPderive}
i\hbar \partial_t\psi=\frac{\delta E(\psi)}{\delta \overline{\psi}}=\left[-\frac{\hbar^2}{2m}\nabla^2+V(\bx)+Ng|\psi|^2\right]\psi, \quad \bx\in {\Bbb R}^3,
\ \ t>0.
\end{equation}
In most BEC experiments, the trapping potential has been taken as the harmonic oscillator potential \cite{Anderson,Baocai2013,Bradley,Davis,PitaevskiiStringari}
\begin{equation}\label{eq:hp}
V({\bx}) =\frac{m}{2}\left(\omega_x^2 x^2+\omega_y^2 y^2+\omega_z^2 z^2\right),
\qquad \bx=(x,y,z)^T\in {\Bbb R}^3,
\end{equation}
where $\omega_x$, $\omega_y$ and $\omega_z$ are the trap frequencies in $x$-,
$y$- and $z$-direction, respectively. Without loss of generality, we assume
that $\omega_x\leq\omega_y\leq\omega_z$ throughout the paper. For other
trapping potentials used in BEC experiments,
such as box potential, double-well potential and optical lattice potential,
we refer to \cite{Baocai2013,BlochDZ,Pethick,Pitaevskii,PitaevskiiStringari,Rosen} and references therein.

The derivation of the GPE (\ref{eq:GPderive}) from the linear Schr\"{o}dinger equation
(\ref{eq:Nbdy}) for a BEC (or a system of $N$ identical particles) based on mean field approximation
-- dimension reduction -- was
formally obtained by Pitaevskii \cite{Pitaevskii}  and  Gross \cite{Gross} independently in 1960s.
Since the first experimental observation of BEC in 1995, much attention has been paid
to provide mathematical justification for the derivation when $N$ is large enough:
For ground states, Lieb et al. \cite{LiebSeiringer,LiebSeiringerPra2000}
proved rigorously that the GP energy (\ref{GPenergy})
approximates the  energy of the many-body system correctly in the mean field regime;
and for dynamics,  Yau et al. \cite{Erdos} established the validity of the
GPE  (\ref{eq:GPderive}) as an approximation for  (\ref{eq:Nbdy}), which inspired great
interests in the study on dynamics for  such many body system recently \cite{ChenT,ChenX,Klainer}.
The above GPE (\ref{eq:GPderive})  is a very simple equation, which is very convenient
for mathematical analysis and numerical calculations, and in the case of the BEC alkali gases,
appears to give a rather good quantitative description of the behavior in a large
variety of experiments \cite{Baocai2013,Pethick,Pitaevskii,PitaevskiiStringari}.
It has  become the fundamental mathematical model
for studying theoretically the ground states and dynamics of BECs \cite{Baocai2013,Pethick,Pitaevskii,PitaevskiiStringari}.

\subsection{Nondimensionalization} In order to study theoretically BECs, we
nondimensionalize the GPE (\ref{eq:GPderive}) with the harmonic trapping potential (\ref{eq:hp})
under the normalization (\ref{GPnorm}) and
introduce \cite{Baocai2013,Pethick,Pitaevskii,PitaevskiiStringari}
\begin{equation}
\label{eq:scale}
\tilde{t} =  \frac{t}{t_s}, \quad \tilde{\bx}=\frac{\bx} {x_s},
\quad  \tilde{\psi}\left(\tilde{\bx},\tilde{t}\right) =  x_s^{3/2}\psi
\left({\bx}, t\right), \quad  \tilde{E}(\tilde{\psi}) = \frac{E(\psi)}{E_s},
\end{equation}
where $t_s=\frac{1}{\omega_x}$, $x_s=\sqrt{\frac{\hbar}{m\omega_x}}$ and $E_s=\hbar \omega_x$
are the scaling parameters of dimensionless
time, length and energy units,
respectively.  Plugging (\ref{eq:scale}) into (\ref{eq:GPderive}), multiplying by
$t_s^2/mx_s^{1/2}$, and  then removing all $\tilde{}$,  we obtain  the
following dimensionless GPE under the normalization  (\ref{GPnorm}) in 3D \cite{Baocai2013,Pethick,Pitaevskii,PitaevskiiStringari}:
\begin{equation}
\label{eq:ngpe}
i\partial_t\psi({\bx},t)
=\left[-\frac{1}{2}\nabla^2+V({\bx})
+\kappa|\psi({\bx},t)|^2\right]\psi({\bx},t),\quad \bx\in {\Bbb R}^3,
\ \ t>0,
\end{equation}
where $\kappa=\frac{4\pi N a_s}{x_s}$  is the
dimensionless interaction constant, the  dimensionless trapping potential is given as
\cite{Baocai2013,Pethick,Pitaevskii,PitaevskiiStringari}
\begin{equation}\label{hartrap}
V({\bx}) = \frac{1}{2}\left(x^2+\gamma_y^2y^2+\gamma_z^2z^2\right), \  \bx\in {\Bbb R}^3,
\ \ \hbox{with}\  \gamma_y=\frac{\omega_y}{\omega_x}\ge1,\ \gamma_z=\frac{\omega_z}{\omega_x}\ge1,
\end{equation}
and dimensionless energy functional $E(\psi)$ is defined as
\cite{Baocai2013,Pethick,Pitaevskii,PitaevskiiStringari}
\begin{equation}
\label{eq:denergy}
E(\psi)=\int_{{\Bbb R}^3} \left[ \frac{1}{2}|\btd \psi(\bx,t)|^2 + V(\bx)
 |\psi(\bx,t)|^2 +\frac{\kappa}{2} |\psi(\bx,t)|^4\right]\; d\bx.
\end{equation}

\subsection{Dimension reduction} In many BEC experiments
\cite{Anderson,Bradley,Davis,Pethick,Pitaevskii,PitaevskiiStringari}, the trapping potential
(\ref{hartrap}) is anisotropic, i.e. $\gamma_z\gg1$ and/or $\gamma_y\gg1$,
and then the GPE in 3D can be further reduced to a GPE in
two dimensions (2D) or one dimension (1D). Assume the initial data
for the 3D GPE (\ref{eq:ngpe}) is given as
\begin{equation}
\psi(\bx,0)=\psi_0(\bx),\qquad \bx \in {\mathbb{R}}^3,\label{spsorig3}
\end{equation}
and define the linear operator $H$ as
\begin{equation} \label{Hamin}
H =-\frac{1}{2} \Delta + V(\bx)=-\frac{1}{2} \nabla^2 + V(\bx),\qquad \bx \in {\mathbb{R}}^3.
\end{equation}

When $\gamma_z\gg1$ and $\gamma_y=O(1)$ ($\Leftrightarrow$ $\omega_z\gg\omega_x$
and $\omega_y=O(\omega_x)$), i.e. disk-shaped condensate with strong
confinement in the $z$-direction \cite{Anderson,Bradley,Davis,Pethick,PitaevskiiStringari},
then the linear operator $H$ can be split  as
\begin{equation} \label{Hamin2d1}
H =-\frac{1}{2} \Delta_\perp + V_2(\bx_\perp)-\frac{1}{2}\partial_{zz}+\frac{z^2}{2\varepsilon^4}
:=H_\perp+H_z^\varepsilon=H_\perp+\frac{1}{\varepsilon^2}H_{\tilde{z}},\quad \bx\in{\Bbb R}^3,
\end{equation}
where $\bx_\perp=(x,y)^T\in {\mathbb{ R}}^2$,
$\Delta_\perp=\partial_{xx}+\partial_{yy}$,
$V_2(\bx_\perp)=\frac{1}{2}(x^2+\gamma_y^2 y^2)$,
$H_\perp:=-\frac{1}{2} \Delta_\perp + V_2(\bx_\perp)$, $\varepsilon=1/\sqrt{\gamma_z}$,
$z=\varepsilon\tilde{z}$ and
\begin{equation}\label{Ham345}
H_z^\varepsilon:=-\frac{1}{2}\partial_{zz}+\frac{z^2}{2\varepsilon^4}
=\frac{1}{\varepsilon^2}
\left[-\frac{1}{2}\partial_{\tilde{z}\tilde{z}}+\frac{\tilde{z}^2}{2}
\right]:=\frac{1}{\varepsilon^2}H_{\tilde{z}},  \qquad z,\tilde{z}\in {\Bbb R}.
\end{equation}
For $H_{\tilde{z}}$ in (\ref{Ham345}), we know that the following linear
eigenvalue problem
\begin{equation}\label{eigen1d}
H_{\tilde{z}}\,\chi(\tilde{z})=\left[-\frac{1}{2}\partial_{\tilde{z}\tilde{z}}+
\frac{\tilde{z}^2}{2}\right]\chi(\tilde{z})=\mu\,\chi(\tilde{z}),  \qquad \tilde{z}\in{\mathbb R},
\end{equation}
with $\|\chi\|^2:=\int_{\mathbb{R}}|\chi(\tilde{z})|^2\,d \tilde{z}=1$
admits distinct orthonormal eigenfunctions  $
\chi_k(\tilde{z})$ with corresponding eigenvalues $
\mu_k$ for $k=0,1,\ldots\;$
In fact, they form  an orthonormal basis of
$L^2(\mathbb R)$ and can be chosen as \cite{Baocai2013,BaoJakschP,bamsw,Pethick,Pitaevskii,PitaevskiiStringari}
\begin{equation}\label{eig1d99}
\mu_{{k}} = \frac{k+1}{2},\quad
\chi_{{k}}(\tilde{z})=\frac{1}{\pi^{1/4}\sqrt{2^k\,k!}}e^{-\tilde{z}^2/2}\,H_k(\tilde{z}),
\ \ \tilde{z}\in {\Bbb R}, \quad
k=0,1,2,\ldots\;,
\end{equation}
with $H_k(\tilde{z})$  the standard Hermite polynomial of degree $k$.
Thus  $\left(\chi_{{k}}^{\varepsilon}(z),\mu_{{k}}^{\varepsilon}
\right)$ for $k\ge0$ are orthonormal eigenpairs to the operator $H_z^{\varepsilon}$ with
\begin{equation}\label{eig1d98}
\mu_{{k}}^{\varepsilon} = \frac{\mu_{{k}}}{{\varepsilon^2}}=\frac{k+1}{2\varepsilon^2},\qquad
\chi_{{k}}^{\varepsilon}(z)=\frac{1}{\sqrt{\varepsilon}}\chi_{{k}}(\tilde{z})
=\frac{1}{\sqrt{\varepsilon}}\chi_{{k}}\left(\frac{z}{\varepsilon}\right),\quad
z\in{\Bbb R}.
\end{equation}
For simplicity of notation, here we only consider ``pure state'' case in the strong confinement
direction, especially the ``ground state'' case
\cite{Baocai2013,BaoJakschP,bamsw,Pethick,Pitaevskii,PitaevskiiStringari}. Assuming that the initial data
$\psi_0$ in (\ref{spsorig3}) satisfies
\begin{equation}\label{init3d567}
\psi_0(\bx)\approx \psi_2(\bx_\perp)\chi_{{0}}^{\varepsilon}(z),
\qquad \bx \in {\mathbb{R}}^3,\qquad 0<\varepsilon \ll1,
\end{equation}
noting the scale separation in (\ref{Hamin2d1}), when $\varepsilon \to0^+$, the solution
$\psi$ to the 3D GPE (\ref{eq:ngpe}) can be well approximated as \cite{Baocai2013,BaoJakschP,bamsw,Pethick,Pitaevskii,PitaevskiiStringari}
\begin{equation}
\label{ProjectAppro}
\psi(\bx,t) \approx
\psi_{2}(\bx_\perp,t)\; \chi_{{0}}^{\varepsilon}(z)\;
e^{-i\, \mu_{0}^\varepsilon\, t},
\qquad \bx \in {\mathbb{R}}^3, \quad t\ge0.\end{equation}
Plugging  (\ref{ProjectAppro}) into (\ref{eq:ngpe}) and
then  multiplying by $\chi_{{0}}^{\varepsilon}(z)\,
e^{i\, \mu_{0}^\varepsilon\, t}$, integrating for $z$ over ${\mathbb{R}}$,
 we obtain formally the GPE in 2D with $\psi_2:=\psi_2(\bx_\perp,t)$ as
 \cite{Baocai2013,BaoJakschP,bamsw,Pethick,Pitaevskii,PitaevskiiStringari}
\begin{equation} \label{2DGPE}
i \partial_t \psi_{{2}} = \left[ -\frac{1}{2}\Delta_\perp
 +V_{2}(\bx_\perp) +\kappa\sqrt{\frac{\gamma_z}{2\pi}}\;
 |\psi_2|^2\right]\psi_{2}, \quad
\bx_\perp \in {\mathbb{R}}^2, \ \ t>0.
\end{equation}
The above dimension reduction from 3D to 2D is mathematically and rigorously
justified in the very weak interaction regime \cite{BaoBenCai,bamsw}, i.e.
 $\kappa=O(\varepsilon)=O(1/\sqrt{\gamma_z})$ as $\varepsilon\to0^+$.
 However, for the strong interaction
regime, i.e. $\kappa=O(1)$ and $\varepsilon\to0^+$, it is very challenging.
The key difficulty is due to that the energy associated to the 2D GPE (\ref{2DGPE})
is unbounded in this regime. Recently, by using a proper re-scaling,
the dimension reduction is justified in this regime too \cite{BLM}.

Similarly, when $\gamma_z\gg1$ and $\gamma_y\gg1$ ($\Leftrightarrow$ $\omega_z\gg\omega_x$
and $\omega_y\gg\omega_x$), i.e. cigar-shaped condensate with strong
confinement in the $(y,z)$-plane \cite{Anderson,Bradley,Davis,Pethick,PitaevskiiStringari},
the 3D GPE (\ref{eq:ngpe}) can be
reduced to the following GPE in 1D as  \cite{Baocai2013,BaoJakschP,Pethick,Pitaevskii,PitaevskiiStringari}
\begin{equation} \label{1DGPE}
i \partial_t \psi_{{1}}(x,t) = \left[ -\frac{1}{2}\partial_{xx}
 +\frac{x^2}{2} +\kappa\frac{\sqrt{\gamma_y\gamma_z}}{2\pi}\;
 |\psi_1(x,t)|^2\right]\psi_{1}(x,t), \ \
x \in {\mathbb{R}}, \  t>0.
\end{equation}

Then the 3D GPE (\ref{eq:ngpe}), 2D GPE (\ref{2DGPE}) and 1D GPE (\ref{1DGPE}) can  be
written in a unified way \cite{Baocai2013,BaoJakschP,Pethick,Pitaevskii,PitaevskiiStringari}
 \begin{equation}\label{eq:gpeg}
i\partial_t{\psi(\bx,t)}=\left[-\frac{1}{2}\btd^2 +
V(\bx)
+ \beta\; |\psi(\bx,t)|^2\right]\psi(\bx,t), \qquad \bx\in {\Bbb R}^d,
\quad t>0,
\end{equation}
where $\beta=\kappa$, $\kappa\sqrt{\gamma_z/2\pi}$ and $\kappa\sqrt{\gamma_y\gamma_z}/2\pi$
when $d=3$, $2$ and $1$, respectively,  and
\begin{equation}
\label{eq:dhp:sec1}
V(\bx)=\frac{1}{2}\begin{cases}
x^2,  & d=1, \\
x^2+\gamma_y^2 y^2, & d=2, \\
x^2+\gamma_y^2 y^2+\gamma_z^2 z^2, &d=3,
\end{cases} \qquad \bx\in {\Bbb R}^d.
\end{equation}
This GPE conserves the normalization (or mass)
\begin{equation}
\label{eq:norm:sec2}
N(\psi(\cdot,t)) =  \int_{\Bbb R^d}|\psi({\bx},t)|^2\; d{\bx} \equiv
\int_{\Bbb R^d}|\psi({\bx},0)|^2\;d{\bx}=1, \qquad t\geq0,
\end{equation}
and the energy per particle
\begin{equation}
\label{eq:energy:sec2}
E(\psi(\cdot,t)) =  \int_{\Bbb R^d}\left[\frac{1}{2}|\nabla  \psi|^2+V({\bx})|\psi|^2
+\frac{\beta}{2}|\psi|^4\right]d{\bx}\equiv  E(\psi(\cdot,0)),\quad t\geq 0.
\end{equation}
In fact, the energy functional  $E(\psi)$ can be split into three
parts as $E(\psi)  = E_{\rm kin}(\psi)+E_{\rm pot}(\psi)
+E_{\rm int}(\psi)$ with the kinetic energy  $E_{\rm kin}(\psi)$, potential energy
$E_{\rm pot}(\psi)$ and  interaction energy $E_{\rm int}(\psi)$ defined as
\begin{equation*}
E_{\rm kin}(\psi) =  \int_{\Bbb R^d}\frac{1}{2}|\nabla\psi|^2d{\bx},\quad
E_{\rm int}(\psi) =  \int_{\Bbb R^d}\frac{\beta}{2}|\psi|^4d{\bx},
\quad E_{\rm pot}(\psi) = \int_{\Bbb R^d}V({\bx})|\psi|^2d{\bx}.
\end{equation*}

\section{Ground states}
To find the stationary state of the GPE (\ref{eq:gpeg}) for a BEC, we write \cite{Baocai2013,BaoDu,Pethick,Pitaevskii,PitaevskiiStringari}
\begin{equation}
\label{eq:state-az:sec2}
\psi({\bx},t)=\phi({\bx})\; e^{-i\mu t}, \qquad \bx\in{\Bbb R}^d, \quad t\ge0,
\end{equation}
where $\mu$ is the chemical potential of the condensate and
$\phi({\bx})$ is a function independent of time. Substituting
(\ref{eq:state-az:sec2}) into (\ref{eq:gpeg}) gives the following
for  $(\mu, \phi)$:
\begin{equation}
\label{eq:charactereq:sec2}
\mu\;\phi({\bx}) =  -\frac{1}{2}\btd^2\phi({\bx})+V({\bx})\phi({\bx})
+\beta|  \phi({\bx})|^2\phi({\bx}), \qquad {\bx}\in \Bbb R^d,
\end{equation}
under the normalization condition
\begin{equation}
\label{eq:gpenorm:sec2}
\|\phi\|^2:=\int_{\Bbb R^d}|\phi({\bx})|^2d{\bx}=1.
\end{equation}
This is a  nonlinear eigenvalue problem with a constraint and any
eigenvalue  $\mu$ can be computed from its corresponding eigenfunction
$\phi({\bx})$ by \cite{Baocai2013,BaoDu,Pethick,Pitaevskii,PitaevskiiStringari}
\begin{equation}
\label{eq:mu-energy:sec2}
\mu=\mu(\phi)
= E(\phi)+\int_{\Bbb R^d}\frac{\beta}{2}|\phi({\bx})|^4d{\bx}=E(\phi)
+E_{\rm  int}(\phi).
\end{equation}

The ground state of a BEC is usually defined as the minimizer of  the
following nonconvex (or constrained) minimization problem \cite{Baocai2013,BaoDu}:
Find $ \phi_g\in S$ such that
\begin{equation}
\label{eq:minp:sec2}
E_g:=E(\phi_g) = \min_{\phi\in S} E(\phi), \qquad \hbox{with}\
\mu_g:=\mu(\phi_g)=E(\phi_g)+E_{\rm  int}(\phi_g),
\end{equation}
where $S=\{\phi \ |\ \|\phi\|=1, \ E(\phi)<\infty\}$ is the  unit sphere and
$\mu_g$ is the corresponding chemical potential.
 It is easy to show that the ground state $\phi_g$ is an  eigenfunction
of the nonlinear eigenvalue problem (\ref{eq:charactereq:sec2})
under the constraint (\ref{eq:gpenorm:sec2}), which is the Euler-Lagrangian equation of
constrained minimization problem (\ref{eq:minp:sec2}).
Any eigenfunction of
(\ref{eq:charactereq:sec2}) whose energy is larger than that  of the ground
state is usually called  excited states in the  physics literatures.

\subsection{Existence and uniqueness}
Denote the best Sobolev constant $C_b$ in 2D as
\begin{equation}\label{eq:bestcons:2d}
C_{b}:=\inf_{0\neq f\in H^1(\Bbb R^2)}\frac{\|\nabla f\|_{L^2(\Bbb R^2)}^2\,
\|f\|_{L^2(\Bbb R^2)}^2}{\|f\|_{L^4(\Bbb R^2)}^4}.
\end{equation}
The best constant $C_b$ can be attained at
some $H^1$ function \cite{Baocai2013} and it is crucial in
considering the existence of ground states in 2D.
For existence and uniqueness of the ground state to (\ref{eq:minp:sec2}), we have the following results.

\begin{theorem}[Existence and uniqueness \cite{Baocai2013,LiebSeiringerPra2000}]
Suppose  $V(\bx)\ge 0$ ($\bx\in\Bbb R^d$) in the energy
functional (\ref{eq:energy:sec2}) satisfies the confining condition
$\lim\limits_{|\bx|\to\infty}V(\bx)=\infty$,
then there exists a ground state $\phi_g\in S$ for (\ref{eq:minp:sec2}) if one of the following holds:
(i)  $d=3$,  $\beta\ge0$;
(ii)  $d=2$,  $\beta>-C_{b}$;
(iii)  $d=1$, for all $\beta\in \Bbb R$.
Moreover, the ground state  can be chosen as nonnegative $|\phi_g|$, and
$\phi_g(\bx)=e^{i\theta_0}|\phi_g(\bx)|$ for some constant  $\theta_0\in\Bbb R$.
For $\beta\ge0$, the nonnegative ground state $\phi_g$ is unique.
If the potential $V(\bx)\in L_{\rm loc}^2$, the nonnegative ground state is strictly positive.
In contrast, there exists no ground state if one of the following holds:
(i$^\prime$)  $d=3$,  $\beta<0$;
(ii$^\prime$)  $d=2$,  $\beta\leq-C_{b}$.
\end{theorem}

For the ground state $\phi_g\in S$ of (\ref{eq:minp:sec2})
with the harmonic potential (\ref{eq:dhp:sec1}), we have the following properties.

 \begin{theorem}[Virial identity \cite{Baocai2013,PitaevskiiStringari}]  The ground state
 $\phi_g\in S$ of (\ref{eq:minp:sec2}) satisfies the following virial identity
 \begin{equation}\label{eq:virial:sec2}
 2E_{\rm kin}(\phi_g)-2\;E_{\rm pot}(\phi_g)+d\;E_{\rm int}(\phi_g)=0.
 \end{equation}
\end{theorem}

\begin{theorem}[Symmetry \cite{Baocai2013,LiebSeiringerPra2000}] Suppose $\gamma_y=\gamma_z=1$ in (\ref{eq:dhp:sec1}),
i.e. the harmonic trapping potential $V(\bx)$ is radially/spherically
symmetric in 2D/3D and monotone increasing,
then the positive ground state  $\phi_g\in S$ of (\ref{eq:minp:sec2})
must be radially/spherically symmetric in 2D/3D
and monotonically decreasing, i.e. $\phi_g(\bx)=\phi_g(r)$ with $r=|\bx|$ for
$\bx\in {\Bbb R}^d$.
 \end{theorem}

\begin{theorem} [Decay at far-field \cite{Baocai2013}]
When $\beta\ge0$, for any $\nu>0$,  there exists a constant $C_{\nu}>0$ such that
\begin{equation}\label{eq:expon:sec2}
|\phi_g(\bx)|\leq C_{\nu}\;e^{-\nu|\bx|},\quad \bx\in\Bbb R^d, \qquad  d=1,2,3.
\end{equation}
\end{theorem}

\subsection{Approximations under the harmonic
potential} For any fixed $\beta\ge0$ in (\ref{eq:energy:sec2}),
we denote the positive ground state of
(\ref{eq:minp:sec2}) with (\ref{eq:dhp:sec1}) as $\phi_g:=\phi_g^\beta$
and the corresponding energy and chemical potential as
$E_g:=E_g^\beta=E(\phi_g^\beta)$ and $\mu_g:=\mu_g^\beta=\mu(\phi_g^\beta)$, respectively.
When $\beta=0$, i.e. linear case,
the exact ground state $\phi_g^0$ can be found as \cite{Baocai2013,BaoDu,Pethick,Pitaevskii,PitaevskiiStringari}
\begin{equation*}
E_g^0=\mu_g^0=\frac{1}{2}\begin{cases}
1, \\
1+\gamma_y, \\
1+\gamma_y+\gamma_z,
\end{cases}
\phi_g^0(\bx)=\begin{cases}
\frac{1}{\pi^{1/4}} \;
e^{-x^2/2}, &d=1,\\
\frac{\gamma_y^{1/4}}{\pi^{1/2}} \;
e^{-(x^2+\gamma_y y^2)/2}, &d=2,\\
\frac{(\gamma_y \gamma_z)^{1/4}}{\pi^{3/4}} \;
e^{-(x^2+\gamma_y y^2+\gamma_z z^2)/2}, &d=3.
\end{cases}
\end{equation*}
When $|\beta|=o(1)$ in (\ref{eq:energy:sec2}), i.e. weak interaction case,
the ground state $\phi_g^\beta$ can be approximated by $\phi_g^\beta(\bx)\approx \phi_g^0(\bx)$
for $\bx\in {\Bbb R}^d$, and the corresponding energy $E_g^\beta$ and chemical potential $\mu_g^\beta$
 can be approximated  with $C_d=\int_{{\Bbb R}^d} |\phi_g^0(\bx)|^4\,d\bx$ as
\begin{equation*}
E_g^\beta\approx E(\phi_g^0)=E_g^0+\frac{\beta}{2}C_d=E_g^0+O(\beta), \ \
\mu_g^\beta\approx \mu(\phi_g^0)=\mu_g^0+\beta C_d=\mu_g^0+O(\beta),
\end{equation*}
where $C_1=\sqrt{\pi/2}$, $C_2=\sqrt{\gamma_y}/2\pi$ and
$C_3=\sqrt{\gamma_y\gamma_z}/(2\pi)^{3/2}$.

When $\beta\gg1$, the ground state $\phi_g^\beta$ can be well approximated by the Thomas-Fermi (TF)
approximation $\phi_g^\beta\approx \phi_g^{\rm TF}$ \cite{Baocai2013,PitaevskiiStringari}, i.e.
by dropping the diffusion term (e.g. the first term
on the right hand side of (\ref{eq:charactereq:sec2})), we obtain
 \begin{equation} \label{eq:gpegs:sec7}
\mu_g^{\rm TF} \phi_g^{\rm TF}(\bx)= V(\bx)\phi_g^{\rm TF}(\bx)+
\beta |\phi_g^{\rm TF}(\bx)|^2\phi_g^{\rm TF}(\bx), \qquad
\bx\in {\Bbb R}^d,
 \end{equation}
with $\mu_g^{\rm TF}\approx \mu_g^\beta$. Solving the above equation, we get
 \begin{equation} \label{eq:gss:sec7}
  \phi_g^\beta(\bx)\approx \phi_g^{\rm TF}(\bx)=\begin{cases}
\sqrt{\left(\mu_g^{\rm TF} -V(\bx)\right)/\beta},
&\ V(\bx)< \mu_g^{\rm TF},\\
0, & \hbox{otherwise},
\end{cases}
 \end{equation}
where $\mu_g^{\rm TF}$ is chosen to satisfy the normalization $\|\phi_g^{\rm TF}\|=1$,
which can be computed as \cite{Baocai2013,BaoDu,Pethick,Pitaevskii,PitaevskiiStringari}
\[
\mu_g^\beta\approx \mu_g^{\rm TF}=\begin{cases}
 \frac{1}{2}\left(\frac{3\beta}
{2}\right)^{2/3},\\
\left(\frac{\beta \gamma_y}{\pi}\right)^{1/2},\\
\frac{1}{2}\left(\frac{15\beta \gamma_y\gamma_z}{4\pi}\right)^{2/5},
\end{cases}
\
E_g^\beta\approx E_g^{\rm TF}=\begin{cases}
\frac{3}{10}\left(\frac{3\beta}{2}
\right)^{2/3}, &d=1,\\
\frac{2}{3}\left(\frac{\beta \gamma_y}{\pi}\right)^{1/2}, &d=2,\\
\frac{5}{14}\left(\frac{15\beta \gamma_y\gamma_z}{4\pi}\right)^{2/5}, &d=3,
\end{cases}
 \]
with $E_g^{\rm TF}:=\mu_g^{\rm TF}-E_{\rm  int}(\phi_g^{\rm TF})$.
For fixed $\gamma_y\ge1$ and $\gamma_z\ge1$ in (\ref{eq:dhp:sec1})
and when $\beta\gg1$ (e.g. $N\gg1$),
from the above TF approximation,  we can get the typical lengthes (i.e.
$R_x^{\rm TF}=\sqrt{2\mu_g^{\rm TF}}$, $R_y^{\rm TF}=\sqrt{2\mu_g^{\rm TF}}/\gamma_y$
and $R_z^{\rm TF}=\sqrt{2\mu_g^{\rm TF}}/\gamma_z$ of the support of the TF approximation $\phi_g^{\rm TF}$
in  $x$-, $y$- and $z$-directions, respectively )
-- TF radius-- of the ground state $\phi_g^\beta$ for a BEC as:
$R_x^{\rm TF}=O(\beta^{1/(d+2)})=O(N^{1/(d+2)})$ for $d=1,2,3$,
$R_y^{\rm TF}=O(\beta^{1/(d+2)})=O(N^{1/(d+2)})$ for $d=2,3$, and
$R_z^{\rm TF}=O(\beta^{1/5})=O(N^{1/5})$ for $d=3$.
In addition,  we also have $E_g^\beta\approx E_g^{\rm TF}=\frac{d+2}{d+4}\mu_g^{\rm TF}\approx
\frac{d+2}{d+4}\mu_g^\beta=O(\beta^{2/(d+2)})=O(N^{2/(d+2)})$,
$\|\phi_g^\beta\|_{L^\infty}\approx\phi_g^{\rm TF}({\bf 0})=O(\beta^{-d/2(d+2)})=O(N^{-d/2(d+2)})$
for $d=1,2,3$.
Thus it is easy to see that there is
no limit of the ground state $\phi_g^\beta$ when $\beta\to\infty$
under the standard physics scaling (\ref{eq:scale}) for a BEC.
In addition, for computing the ground states and dynamics of a BEC,
the bounded computational domain needs to be chosen depending on $\beta$
such that the truncation error can be negligible!

\subsection{Numerical methods}
Various numerical methods for computing the
ground state $\phi_g$ in (\ref{eq:minp:sec2}) have been
proposed and studied in the literature \cite{Baocai2013,BaoChernLim,BaoDu,BaoWang2,Cerim,Ming}.
Among them, one of the most efficient and simple methods is
the following {\sl gradient flow with discrete normalization} (GFDN) \cite{Baocai2013,BaoDu}.
Choose a time step $\tau:=\Delta t>0$ and denote time steps as
$t_n=n\tau$ for $n=0,1,\ldots$ At each time interval $[t_n,t_{n+1})$,
by applying the steepest decent method to the energy functional
$E(\phi)$ without constraint  and then projecting
the solution back to the unit sphere $S$ at $t=t_{n+1}$
so as to satisfy the constraint (\ref{eq:gpenorm:sec2}), we have
\begin{align}
\label{eq:ngf1:sec3}
&\partial_t\phi = -\frac{1}{2}\frac{\delta E(\phi)}{\delta \phi}=
\left[\frac{1}{2}\nabla^2 - V(\bx) -\beta\, |\phi|^2\right]\phi, \quad
 t_n<t<t_{n+1},   \\
\label{eq:ngf2:sec3}
&\phi(\bx,t_{n+1})\stackrel{\triangle}{=}
\phi(\bx,t_{n+1}^+)=\frac{\phi(\bx,t_{n+1}^-)}{\|\phi(\cdot,t_{n+1}^-)\|},
\qquad \bx\in {\Bbb R}^d, \quad n\ge 0,
\end{align}
where $\phi:=\phi(\bx,t)$, $\phi(\bx,t_{n}^\pm)=\lim_{t\to t_{n}^\pm}\phi(\bx,t)$, and with the initial data
\begin{equation}
\label{eq:ngf3:sec3}
\phi(\bx,0)=\phi_0(\bx),\quad \bx\in {\Bbb R}^d.
\end{equation}
In fact, the gradient flow  (\ref{eq:ngf1:sec3}) can be obtained from the GPE (\ref{eq:gpeg})
by $t\to -it$, thus the GFDN is known as {\sl imaginary time method}
in physics literatures \cite{Cerim,Ming}.

For the above GFDN, suppose $V(\bx)\ge0$ for $\bx \in {\Bbb R}^d$ and $\|\phi_0\|^2:=\int_{{\Bbb R}^d}|\phi_0(\bx)|^2\,d\bx$ $=1$, then we have \cite{Baocai2013,BaoDu}

\begin{theorem}[Energy diminishing \cite{BaoDu}] For $\beta=0$,
the GFDN (\ref{eq:ngf1:sec3})-(\ref{eq:ngf3:sec3})
is energy diminishing for any time step $\tau>0$ and
initial data $\phi_0$, i.e.
\begin{equation}
\label{eq:dphi:sec3}
E(\phi(\cdot,t_{n+1}))\le E(\phi(\cdot, t_n))\le \cdots
\le E(\phi(\cdot,0))=E(\phi_0), \;\; n=0,1,2,\cdots.
\end{equation}
\end{theorem}

Let $\tau\to0$ in (\ref{eq:ngf1:sec3})-(\ref{eq:ngf3:sec3}),
we can obtain the following {\sl normalized gradient flow} (NGF) \cite{BaoDu}
\begin{equation}
\label{eq:nkngf1:sec3}
\partial_t\phi(\bx,t) = \left[\frac{1}{2}\nabla^2  - V(\bx) -\beta\, |\phi|^2
+\mu_\phi(t)\right]\phi, \qquad
\bx\in {\Bbb R}^d, \quad  t\ge 0,
\end{equation}
where
\begin{equation}
\mu_\phi(t)=\frac{\mu(\phi(\cdot,t))}{\|\phi(\cdot,t)\|^2}=\frac{1}{\|\phi(\cdot,t)\|^2}
\int_{\Bbb R^d}\left[\frac{1}{2}|\nabla  \phi|^2+V({\bx})|\phi|^2
+\beta|\phi|^4\right]d{\bx}.
\end{equation}

\begin{theorem}[Energy diminishing \cite{BaoDu}]  The NGF
(\ref{eq:nkngf1:sec3}) with (\ref{eq:ngf3:sec3}) is normalization conservative and
energy diminishing, i.e.
\begin{equation}\label{eq:ncphi:sec3}
\|\phi(\cdot,t)\|\equiv\|\phi_0\|=1, \quad
\frac{d}{d t}E(\phi)=- 2\left\|\partial_t\phi(\cdot,t)\right\|^2\le 0\;,
\qquad t\ge0,
\end{equation}
which in turn implies
\begin{equation}
E(\phi(\cdot, t))\ge E(\phi(\cdot,s)), \qquad 0\le t\le s<\infty.
\end{equation}
\end{theorem}

With the above two theorems, the positive ground state can be obtained
from the GFDN as $\phi_g(\bx)=\lim_{t\to \infty} \phi(\bx,t)$ provided
that $\phi_0$ is chosen as a positive function and time step $\tau$ is not too big
when $\beta\ge0$ \cite{Baocai2013,BaoDu}. In addition, the GFDN (\ref{eq:ngf1:sec3})-(\ref{eq:ngf3:sec3})
can be discretized by the {\sl backward Euler finite difference}
(BEFD) discretization \cite{Baocai2013,BaoDu}.  For simplicity of
notation, here we only present the BEFD for the GFDN in 1D truncated on
a bounded interval $U=(a,b)$ with
homogeneous Dirichlet boundary conditions.
Choose a mesh size $h:=\Delta x=(b-a)/M>0$ with $M$ a positive integer,
denote grid points as $x_j=a+jh$ for $j=0,1,\ldots,M$, and
let $\phi^{n}_j$ be the numerical approximation of $\phi(x_j,t_n)$.
Then a BEFD discretization for the GFDN in 1D reads \cite{Baocai2013,BaoDu}
\begin{equation*}
\begin{split}
&\frac{\phi_j^{(1)}-\phi_j^n}{\tau}=\frac{\phi_{j+1}^{(1)}-2\phi_{j}^{(1)}+\phi_{j-1}^{(1)}}{2h^2}
-\left[V(x_j)+\beta \left(\phi_j^{n}\right)^2\right]
\phi_j^{(1)},  \quad 1\le j\le M-1,\\
&\phi_0^{(1)}=\phi_M^{(1)}=0,\quad
\phi_j^0= \phi_0(x_j), \quad \phi_j^{n+1}=\frac{\phi_j^{(1)}}{\|\phi^{(1)}\|_h}, \quad 0\le j\le M,
\quad n\ge0,
\end{split}
\end{equation*}
where $\|\phi^{(1)}\|_h^2:=h\sum_{j=1}^{M-1} |\phi_j^{(1)}|^2$.
This BEFD method is  implicit  and unconditionally stable,
the discretized system can be solved by the Thomas' algorithm,
the memory cost is $O(M)$ and computational cost is $O(M)$ per time step.
The ground state can be obtained numerically from the above BEFD when
$\max\limits_{0\le j\le M} \frac{|\phi_j^{n+1}-\phi_j^n|}{\tau}\le \vep$ with
$\vep$ small enough, e.g. $10^{-6}$.
For extensions to 2D and 3D as well as other numerical methods,
we refer \cite{Baocai2013,BaoChernLim,BaoDu,BaoWang2,Cerim,Ming} and references therein.

\section{Dynamics}
For studying the dynamics of the GPE (\ref{eq:gpeg}), the initial data is usually chosen as
\begin{equation}\label{initial}
\psi(\bx,0)=\psi_0(\bx), \qquad \bx\in{\Bbb R}^d.
\end{equation}
The GPE (\ref{eq:gpeg}) is a dispersive PDE and it is {\sl time reversible or symmetric}, i.e.
it is unchanged under the change of variable in time as $t \rightarrow -t$ and taken conjugate in the equation.
Another important property is {\sl time transverse or gauge invariant}, i.e.
if $V \rightarrow V+\alpha$ with $\alpha$ a given real constant,
then the solution $\psi \rightarrow \psi e^{-i\alpha t}$ which immediately implies that
the density $\rho=|\psi|^2$ is unchanged. It conserves the normalization (or mass) and
energy (or Hamiltonian), i.e.
$N(\psi(\cdot,t))\equiv N(\psi_0)$ and $E(\psi(\cdot,t))\equiv E(\psi_0)$ for $t\ge0$.

\subsection{Well-posedness and dynamical properties}
For studying well-posedness of the GPE (\ref{eq:gpeg}), we introduce the functional spaces
\[
L_V(\Bbb R^d)=\left\{\phi|\int_{\Bbb R^d}V(\bx)|\phi(\bx)|^2d\bx<\infty\right\},\quad
X:=X(\Bbb R^d)=H^1(\Bbb R^d)\cap L_V(\Bbb R^d).
\]

\begin{theorem}[Well-posedness \cite{Baocai2013}] Suppose the trapping
potential is nonnegative and
at most quadratic growth in far field, i.e.,
$V(\bx)\in C^\infty(\Bbb R^d)$ and
$D^{{\bf k}} V(\bx)\in L^\infty(\Bbb R^d)$ for all
${\bf k}\in{\Bbb N}_0^d$ with  $|{\bf k}|\ge 2$, then we have

(i) For any initial data $\psi_0\in X(\Bbb R^d)$,
 there exists a time
$T_{{\rm max}}\in(0,+\infty]$ such that the Cauchy problem of
the GPE (\ref{eq:gpeg}) with (\ref{initial})
 has a unique maximal solution
$\psi\in C\left([0,T_{{\rm max}}),X\right)$. It is maximal in
the sense that if $T_{{\rm max}}<\infty$, then
$\|\psi(\cdot,t)\|_{X}\to\infty$ when  $t\to T^-_{{\rm
max}}$.

(ii) As long as the solution $\psi(\bx,t)$ remains in the energy
space $X$, the {\sl $L^2$-norm} $\|\psi(\cdot,t)\|_2$ and {\sl
energy} $E(\psi(\cdot,t))$  are conserved for
$t\in[0,T_{\rm max})$.

(iii) The solution of the Cauchy problem  is global in time, i.e.,
  $T_{\rm max}=\infty$, if $d=1$ or $d=2$ with $\beta>C_b/\|\psi_0\|_2^2$ or $d=3$ with $\beta\ge0$.
\end{theorem}

\begin{theorem}[Finite time blow-up \cite{Baocai2013}] In 2D and 3D, assume $V(\bx)$ is
at most quadratic growth in far field and satisfies $V(\bx)d+ \bx\cdot \nabla V(\bx)\ge0$ for
$\bx\in{\Bbb R}^d$ ($d=2,3$). When $\beta<0$, for any initial data
$\psi_0(\bx)\in X$ with finite variance $\int_{\Bbb R^d}|\bx|^2|\psi_0|^2\,d\bx<\infty$,
the Cauchy problem of the GPE (\ref{eq:gpeg}) with (\ref{initial})
will blow-up at finite time, i.e. $T_{\rm max}<\infty$, if one of the following holds:
(i) $E(\psi_0)<0$;
(ii) $E(\psi_0)=0$ and ${\rm Im}\left(\int_{\Bbb
R^d}\overline{\psi}_0(\bx)\ (\bx\cdot\nabla\psi_0(\bx))\,d\bx\right)<0$;
(iii) $E(\psi_0)>0$ and ${\rm Im}\left(\int_{\Bbb R^d}
\overline{\psi}_0(\bx)\ (\bx\cdot\nabla\psi_0(\bx))\,d\bx\right)
<-\sqrt{E(\psi_0)d}\;\|\bx\psi_0\|_{L^2}$.
\end{theorem}

If there is no external potential in the GPE (\ref{eq:gpeg}), i.e.  $V(\bx)\equiv0$,
then the momentum and angular momentum are also conserved \cite{Ant,Baocai2013,Sulem}.
The GPE (\ref{eq:gpeg}) admits the plane wave solution as
$\psi(\bx,t) = A e^{i(\mathbf{k} \cdot \bx - \omega t)}$, where
the time frequency $\omega$, amplitude $A$ and spatial wave number $\mathbf{k}$ satisfy the following
{\sl dispersion relation} \cite{Ant,Baocai2013,Sulem}:
$\omega = \frac{|\mathbf{k}|^{2}}{2}+\beta |A|^{2}$.
In 1D, i.e. $d=1$,  when $\beta <0$, it
admits the well-known bright soliton solution as \cite{Ant,Sulem}
\begin{equation}
\label{bright_sol}
\psi_{B}(x,t)=\frac{A}{\sqrt{-\beta}} \textrm{sech}(A(x-vt-x_{0}))
e^{i(vx -\frac{1}{2}(v^{2}-A^{2})t + \theta_{0})},
\qquad x \in {\Bbb R}, \quad t \geq 0,
\end{equation}
where $\frac{A}{\sqrt{-\beta}}$ is the amplitude of the soliton with $A$ a positive real
constant, $v$ is the velocity of the soliton, $x_{0}$
and $\theta_0$ are the initial shifts in space and phase, respectively.
Since the soliton solution is exponentially decaying
 for $|x| \rightarrow + \infty$, then the mass and energy are well defined and given by: $N(\psi_{B})=-\frac{2A}{\beta}$
 and $E({\psi_{B}})=\frac{Av^{2}}{-\beta}+\frac{A^{3}}{-3\beta}$.
When $\beta>0$, it admits dark solitons \cite{PitaevskiiStringari,Sulem}.

Let $\psi:=\psi(\bx,t)$ be the solution of
the GPE (\ref{eq:gpeg}) with the harmonic potential (\ref{eq:dhp:sec1})
and initial data (\ref{initial}) satisfying $\|\psi_0\|=1$, define the center-of-mass
${\bx}_c(t)=\int_{\Bbb R^d}\bx |\psi(\bx,t)|^2\,d\bx$,
square of the condensate width $\delta_\alpha(t)=
\int_{\mathbb{R}^d}\alpha^2|\psi({\bx},t)|^2d{\bx}$ with $\alpha=x$, $y$ or $z$,
and angular momentum expectation $\langle L_z\rangle(t)=
\int_{\mathbb{R}^d} \overline{\psi(\bx,t)} L_z\psi(\bx,t)\,d\bx$ with
$L_z=-i\left(x\partial_y -y\partial_x\right)$ when $d=2,3$.
Then we have \cite{Baocai2013,BaoDuZhang}

\begin{lemma}[Angular momentum expectation \cite{Baocai2013,BaoDuZhang}]
For any initial data $\psi_0(\bx)$ in (\ref{initial}),
when $\gamma_y=1$ in  (\ref{eq:dhp:sec1}), i.e. the trapping potential
is radially/cylindrically symmetric in 2D/3D, then the angular momentum expectation is conserved, i.e.
\begin{equation}
\langle L_z\rangle(t)\equiv \langle L_z\rangle(0)=
\int_{\mathbb{R}^d} \overline{\psi_0(\bx)} L_z\psi_0(\bx)\,d\bx, \qquad t\ge0.
\end{equation}
\end{lemma}

\begin{lemma}[Condensate width \cite{Baocai2013,BaoDuZhang}]
For any initial data $\psi_0(\bx)$ in (\ref{initial}), in 1D without interaction,
i.e. $d=1$ and $\beta=0$ in (\ref{eq:gpeg}),  we have
\begin{equation}
\label{eq:solution_dt_r:sec27}
\dt_x(t) =E(\psi_0)+\left(\delta_x^{(0)}-E(\psi_0)\right)
\cos(2t)+\dt_x^{(1)}\sin(2 t),\quad t\ge0;
\end{equation}
and in 2D with a radially symmetric trap, i.e. $d = 2$ and
$\gamma_y=1$ in (\ref{eq:dhp:sec1}),  we have
\begin{equation}
\label{eq:solution_dt_r:sec28}
\dt_r(t) = E(\psi_0)+\left(\dt_r^{(0)}-E(\psi_0)\right)
\cos(2t)+\dt_r^{(1)}\sin(2 t), \quad t\ge0,
\end{equation}
where $\dt_r(t)= \dt_x(t)+\dt_y(t)$,
$\dt_r^{(0)}:=\dt_x^{(0)}+\dt_y^{(0)}$, and
$\dt_r^{(1)}:=\dt_x^{(1)}+\dt_y^{(1)}$ with
$\dt_{\alpha}^{(0)} = \int_{\mathbb{R}^d}\alpha^2|\psi_0({\bx})|^2
d{\bx}$ and $\dt_{\alpha}^{(1)} = 2\int_{\mathbb{R}^d}\alpha\, {{\rm Im}}
\left(\overline{\psi}_0\partial_\alpha\psi_0\right)\; d{\bx}$ for $\alpha = x$ or $y$.
Thus $\dt_x$ in 1D and $\dt_r$ in 2D are periodic functions with frequency doubling the
trapping frequency.
\end{lemma}

\begin{lemma}[Center-of-mass \cite{Baocai2013,BaoDuZhang,Bao2013}]
For any initial data $\psi_0(\bx)$ in (\ref{initial}), the dynamics of the center-of-mass satisfies
the following second-order ODE
\begin{equation}\label{eq:mc_ODE2:sec2}
\ddot{\bx}_{c}(t)+\Lambda\,{\bx}_c(t)=0,\quad t\ge0,
\end{equation}
with the following initial data
\[
{\bx}_{c}(0) = {\bx}_{c}^{(0)} = \int_{\mathbb{R}^d}\bx|\psi_0({\bx})|^2d{\bx}, \quad
\dot{\bx}_c(0) = {\bx}_{c}^{(1)}=\int_{\Bbb R^d}{\rm Im}(\overline{\psi}_0\nabla \psi_0)\,d\bx,
\]
where $\Lambda$ is a $d\times d$ diagonal matrix as $\Lambda=1$ when $d=1$,
$\Lambda={\rm diag}(1,\gamma_y^2)$ when $d=2$, and
$\Lambda={\rm diag}(1,\gamma_y^2,\gamma_z^2)$ when $d=3$.
This implies that each
component of $\bx_c$ is a periodic function whose frequency is the
same as the trapping frequency in that direction.
\end{lemma}

\begin{lemma}[Exact solution \cite{Baocai2013,BaoDuZhang}]
If the initial data $\psi_0(\bx)$ in (\ref{initial})
is chosen as
\begin{equation}
\label{eq:init5:sec2}
\psi_0(\bx)=\phi_e(\bx-\bx_0)\,e^{i({\bf w}_0\cdot \bx+g_0)}, \qquad
\bx \in {\mathbb R}^d,
\end{equation}
where $\bx_0,{\bf w}_0\in {\mathbb R}^d$ and $g_0\in {\Bbb R}$ are given constants,
and $(\mu_e,\phi_e)$ is a solution of the nonlinear eigenvalue problem (\ref{eq:charactereq:sec2}) with
the constraint (\ref{eq:gpenorm:sec2}), then the
GPE (\ref{eq:gpeg}) with (\ref{eq:dhp:sec1}) and (\ref{eq:init5:sec2})
admits the following unique  exact solution
\begin{equation}
\label{eq:exacts1:sec2}
\psi(\bx,t)=\phi_e(\bx-\bx_c(t))\;e^{-i\mu_e t}\; e^{i({\bf w}(t) \cdot \bx + g(t))},
\qquad \bx\in{\mathbb R}^d, \quad t\ge 0,
\end{equation}
where $\bx_c(t)$ satisfies the second-order ODE (\ref{eq:mc_ODE2:sec2}) with
the initial condition $\bx_c(0)=\bx_0$ and $\dot{\bx}_c(0)={\bf w}_0$, and
${\bf w}(t)$ and $g(t)$ satisfy the following ODEs
\begin{equation}
\dot{\bf w}(t)=-\Lambda\, \bx_c(t), \qquad \dot g(t)
=V(\bx_c(t))=\frac{1}{2}\bx_c(t)\cdot (\Lambda\, \bx_c(t)),
\qquad t>0,
\end{equation}
with initial data ${\bf w}(0)={\bf w}_0$ and $g(0)=g_0$.
\end{lemma}

\subsection{Numerical methods}
Various numerical methods have been
proposed and studied in the literature \cite{Ant,Baocai2013,BaoJakschP,BaoShen,Cerim,Ming} for computing the
dynamics of the GPE (\ref{eq:gpeg}) with (\ref{initial}).
Among them, one of the most efficient and accurate as well as simple methods is
the following {\sl time-splitting sine pseudospectral} (TSSP) method \cite{Ant,Baocai2013,BaoJakschP}.
For simplicity of notation, here we only present the TSSP method for the  GPE (\ref{eq:gpeg})
 in 1D truncated on a bounded interval $U=(a,b)$ with
homogeneous Dirichlet boundary conditions.
Let $\psi^{n}_j$ be the numerical approximation of $\psi(x_j,t_n)$
and $\psi^{n}$ be the solution vector at time $t=t_n=n\tau$ with
components $\{\psi_j^{n}\}_{j=0}^M$,
then a second-order TSSP method for the  GPE (\ref{eq:gpeg})
in 1D reads \cite{Ant,Baocai2013,BaoJakschP}
\begin{equation*}
\begin{split}
&\psi_j^{(1)}=\frac{2}{M}\sum_{l=1}^{M-1}
e^{-i\tau\mu_l^2/4}\;\widetilde{(\psi^n)}_l\;\sin(\mu_l(x_j-a)),\quad
\psi^{(2)}_j=e^{-i\tau(V(x_j)+\beta |\psi_j^{(1)}|^2)}\;\psi_j^{(1)}, \\
&\psi^{n+1}_j=\frac{2}{M}\sum_{l=1}^{M-1}
e^{-i\tau\mu_l^2/4}\;\widetilde{(\psi^{(2)})}_l\;\sin(\mu_l(x_j-a)), \qquad 0\le j\le M,
\end{split}
\end{equation*}
where $\mu_l=l\pi/(b-a)$ for $1\le l\le M-1$ and
$\widetilde{(\psi^n)}_l$ and $\widetilde{(\psi^{(2)})}_l$ are the discrete sine transform (DST)
coefficients of $\psi^n$  and $\psi^{(2)}$, respectively.
This TSSP method for the GPE (\ref{eq:gpeg}) is explicit,
unconditionally stable, second-order accurate in time and spectral-order accurate in space \cite{Ant,Baocai2013,BaoJakschP}.
It is time reversible or symmetric, time transverse invariant, conserves the mass at the discetized
level and has the same dispersive relation as the GPE when $V(\bx)\equiv0$.
The memory cost is $O(M)$ and computational cost  is $O(M\ln M)$ per time step.
For extensions to 2D/3D and other numerical methods,
we refer to \cite{Ant,Baocai2013,BaoJakschP,BaoShen,Cerim,Ming} and references therein.

\subsection{Bogoliubov excitation of ground state}
An important class of time-dependent solutions of the GPE (\ref{eq:gpeg}) is given by the small-amplitude
oscillations, where the changes in space and time of the
wave function (or order parameter) with respect to the stationary states, especially ground states,
are small. In many cases these solutions emphasize the collective behavior exhibited
by the interacting Bose gases and can be interpreted in terms of the elementary excitations of
the system. For describing the dynamics of a BEC, it is natural to consider the linearized
behavior of small perturbations around its ground state $\phi_g$ with chemical potential $\mu_g$
and take the ansatz \cite{Dalfovo,Fetter,GiorginiPS,PitaevskiiStringari}
\begin{equation}\label{BA98}
\psi(\bx,t)=e^{-i\mu_e t}\left[\phi_g(\bx)+u(\bx)e^{-i\omega t}-\overline{v(\bx)}e^{i\omega t}\right],
\qquad \bx\in{\Bbb R}^d, \quad t>0,
\end{equation}
where the Bogoliubov amplitudes $u(\bx)$ and $v(\bx)$ are treated as small and $\omega\in{\Bbb C}$
to be determined. Substituting (\ref{BA98})
into (\ref{eq:gpeg}) and collecting first-order terms proportional to $e^{\pm i\omega t}$, we obtain
the Bogoliubov equations -- linear eigenvalue problem for $(\omega, u,v)$--- as
\cite{Dalfovo,Fetter,GiorginiPS,PitaevskiiStringari}
\begin{equation}\label{BA99}
\begin{split}
&\omega\, u(\bx)= \left[-\frac{1}{2}\nabla^2 +V(\bx)+2\beta |\phi_g(\bx)|^2-\mu_g\right]u(\bx)-\phi_g^2\, v(\bx),
\quad \bx\in{\Bbb R}^d,\\
&-\omega\, v(\bx)= \left[-\frac{1}{2}\nabla^2 +V(\bx)+2\beta |\phi_g(\bx)|^2-\mu_g\right]v(\bx)-
\overline{\phi}_g^2\, u(\bx).
\end{split}
\end{equation}
In many ways, the above Bogoliubov equations are analogous to a nonrelativistic version
of the Dirac equation, with $u$ and $v$ as the particle and hole amplitudes,
including the ($+$, $-$) metric seen in the minus sign on the left hand side
of the second equation compared to the first equation in (\ref{BA99}) \cite{Dalfovo,Fetter,GiorginiPS,PitaevskiiStringari}.
In addition, a detailed analysis shows that physically relevant Bogoliubov eigenfunctions
must satisfy the following {\sl positive} normalization condition
\cite{Dalfovo,Fetter,GiorginiPS,PitaevskiiStringari}:
\begin{equation}
\|u\|^2-\|v\|^2:=\int_{{\Bbb R}^d} \left[|u(\bx)|^2-|v(\bx)|^2\right]\,d\bx=1.
\end{equation}
For solutions of the Bogoliubov equations, especially no external trapping potential in (\ref{eq:gpeg}),
we refer to \cite{Dalfovo,Fetter,GiorginiPS,PitaevskiiStringari} and references therein.

\subsection{Semiclassical scaling and limits}
In the strongly repulsive interaction regime, i.e.
$\beta\gg1$ in the GPE (\ref{eq:gpeg}) with (\ref{eq:dhp:sec1}),
another  scaling  (under the normalization (\ref{eq:norm:sec2}) with $\psi$ being
replaced by $\psi^\vep$) -- semiclassical scaling --  is also very useful in
practice, especially in numerical computation. By choosing
$\bx\to \bx \vep^{-1/2}$ and $\psi = \vep^{d/4}\,\psi^\vep$ with
$0<\vep=1/\beta^{2/(2+d)}<1$ ($\Leftrightarrow t=\frac{1}{\omega_x}$,
$x_s=\sqrt{\hbar/m\vep\omega_x}$ and
$E_s=\hbar\omega_x/\vep$ in (\ref{eq:scale}) for the GPE (\ref{eq:GPderive}) when $d=3$),
we obtain \cite{Baocai2013,BaoJakschP}
\begin{equation}
\label{eq:semiclass:sec7}
i\vep\, \partial_t\psi^\vep(\bx,t)
=\left[-\frac{\vep^2}{2}\nabla^2  + V(\bx)
+ |\psi^\vep(\bx,t)|^2\right]\psi^\vep(\bx,t), \quad {\bx}\in {\Bbb R}^d,\ \  t>0.
\end{equation}
This GPE conserves the following energy
\begin{equation*}
E^\vep(\psi^\vep(\cdot,t))=\int_{{\Bbb R}^d}\left[\frac{\vep^2}{2}|\nabla
\psi^\vep|^2+V(\bx)  |\psi^\vep|^2
+\frac{1}{2}|\psi^\vep|^4\right]d{\bx}\equiv E^\vep(\psi^\vep(\cdot,0)),
\quad t\ge0.
\end{equation*}
Similarly, the nonlinear eigenvalue problem (\ref{eq:charactereq:sec2})
(under the normalization (\ref{eq:gpenorm:sec2}) with $\phi=\phi^\vep$)  reads
\begin{equation}
\label{eq:charac:sec7} \mu^\vep\phi^\vep({\bx}) =
\left[-\frac{\vep^2}{2}\nabla^2
+V({\bx})+|  \phi^\vep(\bx)|^2\right]\phi^\vep(\bx), \quad {\bx}
\in {\Bbb R}^d,
\end{equation}
where the eigenvalue (or chemical potential) $\mu^\vep$
can be computed from its corresponding
eigenfunction $\phi^\vep$ by $\mu^\vep=\mu^\vep(\phi^\vep)=E^\vep(\phi^\vep)+E_{\rm  int}^\vep(\phi^\vep)$
with $E_{\rm  int}^\vep(\phi^\vep)=\frac{1}{2}\int_{\Bbb R^d}|\phi^\vep|^4\,d\bx$.
The constrained minimization problem for ground state collapses to: Find $ \phi_g^\vep\in S$ such that
\begin{equation}
\label{eq:minp:sec3}
E_g^\vep:=E^\vep(\phi_g^\vep) = \min_{\phi^\vep\in S} E^\vep(\phi^\vep), \qquad \hbox{with}\
\mu_g^\vep:=\mu^\vep(\phi_g^\vep)=E^\vep(\phi_g^\vep)+E_{\rm  int}^\vep(\phi_g^\vep).
\end{equation}
Similarly to section 3.2, we can get the TF approximation to the ground state when
$0<\vep\ll 1$:
\begin{equation} \label{eq:gsts:sec7}
\phi_g^\vep(\bx)\approx \phi_g^{\rm TF}(\bx)=\begin{cases}
\sqrt{\mu_g^{\rm TF} -V(\bx)},
&\ V(\bx)< \mu_g^{\rm TF},\\
0, & \hbox{otherwise},
\end{cases}
 \end{equation}
where
\[
\mu_g^\vep\approx \mu_g^{\rm TF}=\begin{cases}
 \frac{1}{2}\left(\frac{3}
{2}\right)^{2/3},\\
\left(\frac{\gamma_y}{\pi}\right)^{1/2},\\
\frac{1}{2}\left(\frac{15\gamma_y\gamma_z}{4\pi}\right)^{2/5},
\end{cases}
\
E_g^\vep\approx E_g^{\rm TF}=\begin{cases}
\frac{3}{10}\left(\frac{3}{2}
\right)^{2/3}, &d=1,\\
\frac{2}{3}\left(\frac{\gamma_y}{\pi}\right)^{1/2}, &d=2,\\
\frac{5}{14}\left(\frac{15\gamma_y\gamma_z}{4\pi}\right)^{2/5}, &d=3.
\end{cases}
 \]
From this TF approximation,
for fixed $\gamma_y\ge1$ and $\gamma_z\ge1$ in (\ref{eq:dhp:sec1})
and when $0< \vep \ll 1$, we have $E_g^\vep\approx E_g^{\rm TF}=\frac{d+2}{d+4}\mu_g^{\rm TF}\approx
\frac{d+2}{d+4}\mu_g^\vep=O(1)$, $\|\phi_g^\vep\|_{L^\infty}\approx\phi_g^{\rm TF}({\bf 0})=O(1)$,
and the TF radius $R_x^{\rm TF}=\sqrt{2\mu_g^{\rm TF}}=O(1)$,
$R_y^{\rm TF}=\sqrt{2\mu_g^{\rm TF}}/\gamma_y=O(1)$
and $R_z^{\rm TF}=\sqrt{2\mu_g^{\rm TF}}/\gamma_z=O(1)$
for $d=1,2,3$. In addition, the ground state $\phi_g^\vep(\bx)$ converges to
$\phi_g^{\rm TF}(\bx)$ uniformly when $\vep\to 0^+$.
Furthermore, for computing numerically the ground states and dynamics of a BEC,
the bounded computational domain can be chosen independent of $\vep$ \cite{Baocai2013,BaoJakschP}.

Taking the WKB ansatz $\psi^\vep(\bx,t)=\sqrt{\rho^\vep(\bx,t)}\,e^{i S^\vep(\bx,t)/\vep}$
with $\rho^\vep=|\psi^\vep|^2$ and $S^\vep$  the density and phase of the wave function, respectively,
inserting it into the GPE (\ref{eq:semiclass:sec7})  and
separating real and imaginary parts,  we obtain
the transport and  Hamilton-Jacobi equations for density and phase, respectively
\cite{Baocai2013,Carles01,GMMP}
\begin{equation}\label{eq:eqHJ:sec7}\begin{split}
&\partial_t\rho^\vep+{\rm div}\, (\rho^\vep\; \btd S^\vep)=0, \qquad \qquad \qquad
\qquad \bx\in{\Bbb R}^d, \quad t>0,\\
&\partial_t S^\vep+\frac{1}{2}|\nabla S^\vep|^2 +
\rho^\vep+V(\bx)
=\frac{\vep^2}{2}\;\frac{1}{\sqrt{\rho^\vep}}\ \Delta \sqrt{\rho^\vep}.
\end{split}
\end{equation}
Furthermore, defining the quantum velocity ${\bf u}^{\vep}=\nabla S^{\vep}$ and current
${\bf J}^{\vep}=\rho^{\vep}\,{\bf u}^{\vep}$, we get from (\ref{eq:eqHJ:sec7}) the Euler system with a
third-order dispersion correction term -- quantum hydrodynamics (QHD) -- as \cite{Baocai2013,Carles01,GMMP}
\begin{equation}
\label{eq:6rho_1:sec7}
\begin{split}
&\partial_t\rho^{\vep}+{\rm div}\,{\bf
J}^{\vep} = 0, \qquad \qquad \qquad
\qquad \bx\in{\Bbb R}^d, \quad t>0,\\
&\partial_t{\bf J}^{\vep}+{\rm div}\left(\frac{{\bf
J}^{\vep}\otimes{\bf
J}^{\vep}}{\rho^{\vep}}\right)+\rho^{\vep}\,\nabla
V(\bx)+\nabla P(\rho^\vep)=\frac{\vep^2}{4}\nabla\left(\rho^{\vep}
\nabla^2\ln\rho^{\vep}\right),
\end{split}
\end{equation}
where the pressure is defined as $P(\rho^\vep)=\left(\rho^{\vep}\right)^2/2$.
Letting $\vep\to0^+$ in (\ref{eq:6rho_1:sec7}), formally we get the Euler
system \cite{Baocai2013,Carles01,GMMP}
\begin{equation}\label{eq:6rho_3:sec7}
\begin{split}
&\partial_t\rho^{0}+{\rm div}\,{\bf
J}^{0} = 0, \qquad \qquad \qquad
\qquad \bx\in{\Bbb R}^d, \quad t>0,\\
&\partial_t{\bf J}^{0}+{\rm div}\left(\frac{{\bf
J}^{0}\otimes{\bf
J}^{0}}{\rho^{0}}\right)+\rho^{0}\,\nabla V(\bx)+\nabla P(\rho^0)=0.
\end{split}
\end{equation}
For mathematical justification of the passage from the GPE
(\ref{eq:semiclass:sec7}) to the Euler system (\ref{eq:6rho_3:sec7}),
we refer to \cite{Baocai2013,Carles01,GMMP} and references therein.

\section{Extensions}
In this section, we will present briefly
mathematical models and theories as well as numerical methods for
rotating BEC based on the GPE with an angular momentum rotation term,
dipolar BEC based on the GPE with
a long-range anistropic dipole-dipole interaction (DDI)  and
spin-orbit-coupled BEC based on coupled GPEs with an internal
atomic Josephon junction (JJ) and an spin-orbit coupling term.

\subsection{For rotating BEC}
At temperatures $T$ much smaller than the critical temperature
$T_c$, following the mean field theory
\cite{Abo,Aftalion,Baocai2013,Caradoc,Fetter,LiebSeiringer1,Matthews,Seiringer},
a BEC in the rotational frame is well described by
the macroscopic wave function $\psi:=\psi(\bx,t)$, whose evolution is
governed by the GPE with an angular momentum rotation term
\begin{equation}\label{eq:gperota:sec5}
i\hbar\partial_t \psi=\left[-\frac{\hbar^2}{2m}\nabla^2 + V(\bx)-\tilde{\Omega} L_z+N
g |\psi|^2\right] \psi,  \qquad \bx\in{\Bbb R}^3,\quad t>0,
\end{equation}
where $\tilde{\Omega}$ is the angular velocity,
$L_z$ is the $z$-component angular momentum operator defined as
$L_z=-i \hbar\left(x\partial_y -y\partial_x\right)$ and $\psi$
satisfies the normalization condition (\ref{GPnorm}).

Under the harmonic potential (\ref{eq:hp}), similarly to
the nondimensionalization in section 2.2 and dimension reduction
in 2.3 from 3D to 2D when $\omega_z\gg \max\{\omega_x,\omega_y\}$ for a disk-shaped condensate
\cite{Aftalion,Baocai2013,BaoDuZhang,BaoWangP}, we can obtain
the following dimensionless GPE with an angular momentum rotation term
in $d$-dimensions ($d=2,3$):
\begin{equation} \label{eq:gpegrot:sec5}
i\;\partial_t \psi=\left[-\frac{1}{2}\nabla^2+
V(\bx) -\Omega L_z + \beta|\psi|^2\right]
\psi, \quad \bx\in {\Bbb R}^d, \quad t>0,
\end{equation}
where $\Omega=\tilde{\Omega}/\omega_x$, $\beta=\kappa$ and
$\kappa\sqrt{\gamma_z/2\pi}$ when $d=3$ and $2$, respectively,
the dimensionless harmonic potential is given in (\ref{eq:dhp:sec1}) for $d=3,2$,
and the dimensionless angular momentum rotation term is given as
$L_z=-i\left(x\partial_y -y\partial_x\right)$. The GPE (\ref{eq:gpegrot:sec5}) conserves the
normalization (\ref{GPnorm}) and energy per particle
\begin{equation*}\label{eq:engd1:sec5}
E(\psi(\cdot,t)) =\int_{{\Bbb
R}^d} \left[\frac{1}{2} \left|\nabla \psi\right|^2+
V(\bx)|\psi|^2-\Omega \overline{\psi}\, L_z \psi +\frac{\beta}{2}\,
|\psi|^4\right]d\bx\equiv E(\psi(\cdot,0)),\ t\ge0.
\end{equation*}

The ground state can be defined the same as
(\ref{eq:minp:sec2}) with the above energy functional.
For the existence and uniqueness as well as nonexistence,
we have \cite{Aftalion,Baocai2013,BaoWangP,Seiringer}

\begin{theorem}[Existence and uniqueness \cite{Aftalion,Baocai2013,BaoWangP,Seiringer}]
Suppose that $V(\bx)$ is taken as the harmonic potential
in (\ref{eq:dhp:sec1}), then we have

i) There exists a ground state of the rotating BEC (\ref{eq:gpegrot:sec5}) when
$|\Omega|<1$ and $\beta\ge0$ in 3D or $\beta>-C_b$ in 2D.

ii) For any $\beta\ge0$,  there exists a critical rotation velocity
$0<\Omega_c^\beta\le 1$ -- {\sl first critical rotation speed} --
depending on $\beta$ such that: when $\Omega_c^\beta<|\Omega|<1$,
quantized vortices will appear in the ground state $\phi_g$.

iii) In 2D with $\gamma_y=1$ (radially symmetric $V(\bx)$),
there exists $\beta_0>0$ such that when $\beta\ge\beta_0$, for $|\Omega|<\Omega_{c_1}^\beta$
($\Omega_{c_1}^\beta$ depends on $\beta$), the ground state  can be chosen as
positive $|\phi_g|$, and $\phi_g(\bx)=e^{i\theta_0}|\phi_g(\bx)|$
for some constant  $\theta_0\in\Bbb R$, and the positive ground
state $\phi_g$ is unique.

iv) There exists no ground state of the rotating BEC (\ref{eq:gpegrot:sec5})
if one of the following holds: (a) $\beta<0$ in 3D or $\beta<-C_b$ in 2D;
 (b) $|\Omega|> 1$.
\end{theorem}
\begin{remark} From the various numerical results,
for radially symmetric $V(\bx)$ in 2D (or cylindrically symmetric in 3D ) and
any fixed $\beta\ge0$, the {\sl first critical rotation speed}
$0<\Omega_c^\beta\le 1$
depends on $\beta$ and:
when $|\Omega|<\Omega_c^\beta$, the ground state  can be chosen as
nonnegative $|\phi_g|$, and $\phi_g(\bx)=e^{i\theta_0}|\phi_g(\bx)|$
for some constant  $\theta_0\in\Bbb R$, and the nonnegative ground
state $\phi_g$ is unique; when $\Omega_c^\beta<|\Omega|<1$,
quantized vortices will appear in the ground state $\phi_g$;
and when $\Omega_c^\beta=|\Omega|$, there exist as least two
different ground states -- one without quantized vortices and one with
quantized vortices. We remark here that
a rigorous mathematical justification is still missing.
\end{remark}
For more results on the ground state of
the rotating BEC (\ref{eq:gpegrot:sec5}) and
efficient and accurate numerical methods  for simulation, such as BEFD \cite{Baocai2013,BaoWangP}
or BEFP \cite{BaoChernLim},
we refer to \cite{Aftalion,Baocai2013,BaoCai2,BaoWangP,Fetter,Seiringer} and references therein.
Similarly, for the well-posedness of the Cauchy problem of (\ref{eq:gpegrot:sec5})
with the initial data (\ref{initial}) and its dynamical properties as well
as efficient and accurate numerical methods, such as TSADI \cite{BaoWang} or TSGLFHP \cite{BaoLiShen},
we refer to \cite{Ant,Baocai2013,Fetter,Seiringer} and references therein. Here we present
a different formulation of the GPE (\ref{eq:gpegrot:sec5}) under the
{\sl rotating Lagrangian coordinates} so that the angular momentum rotation term will be
removed \cite{Bao2013}.

For any time $t\geq 0$, let ${A}(t)$ be an orthogonal rotational matrix  defined as
\begin{equation*}\label{Amatrix}
{A}(t)=\left(\begin{array}{cc}
\cos(\Omega t) &\sin(\Omega t)\\
-\sin(\Omega t) &\cos(\Omega t)
\end{array}\right), \ d =2,
\  {A}(t)=\left(\begin{array}{ccc}
\cos(\Omega t) &\sin(\Omega t) &0 \\
-\sin(\Omega t) &\cos(\Omega t) &0 \\
0 & 0  & 1
\end{array}\right), \  d = 3.
 \end{equation*}
It is easy to verify that $A^{-1}(t) =A^T(t)$ for any $t\ge0$ and ${A}(0) = { I}$ with
 $I$ the identity matrix. For any $t\ge0$, we introduce the {\it rotating Lagrangian coordinates}
$\tbx$ as \cite{Bao2013}
\begin{equation}\label{transform}
\tbx={A}^{-1}(t) \bx=A^T(t)\bx \quad \Leftrightarrow \quad \bx= {A}(t){\tbx},   \qquad \bx\in {\mathbb R}^d,
\end{equation}
and denote the wave function in the new coordinates as $\varphi:=\varphi(\tbx, t)$
\begin{equation}\label{transform79}
\varphi(\tbx, t):=\psi(\bx, t)= \psi\left({A}(t){\tbx},t\right), \qquad \bx\in {\mathbb R}^d, \quad t\geq0.
\end{equation}
Here, we refer the Cartesian coordinates $\bx$ as the {\it Eulerian coordinates}.
Plugging (\ref{transform}) and (\ref{transform79}) into (\ref{eq:gpegrot:sec5}), we obtain the GPE
\begin{equation}\label{GPEvp3}
i\partial_t \varphi(\widetilde{\bx}, t) =
\left[-\frac{1}{2}\nabla^2 + W(\widetilde{\bx}, t) + \beta|\varphi(\widetilde{\bx}, t)|^2
\right]\varphi(\widetilde{\bx}, t), \quad \tbx\in{\mathbb R}^d, \quad t>0,
\end{equation}
where $W(\tbx,t) = V(A(t)\tbx)$ for $\tbx\in{\mathbb R}^d$ and $t>0$,
which is time-independent, i.e. $W(\tbx, t) = V(\tbx)$ if the harmonic potential
(\ref{eq:dhp:sec1}) is radially/cylindically symmetric in 2D/3D, i.e. $\gamma_y=1$.
In addition, the initial data for the GPE (\ref{GPEvp3}) from (\ref{initial}) is
\begin{equation}
\label{Initialgg}
\varphi(\tbx, 0) = \psi(\bx,0)=\psi_0(\bx):=\varphi_0(\bx)=\varphi_0(\tbx),
\qquad \tbx=\bx\in{\Bbb R}^d.
\end{equation}
Based on the above new formulation,
the results and numerical methods developed for nonrotating BEC, such as TSSP \cite{Ant,Baocai2013,BaoJakschP,BaoLiShen,BaoShen},
can be directly applied for analyzing and simulating the dynamics of rotating BEC.

\subsection{For dipolar BEC}
At temperature $T$ much smaller than the critical temperature $T_c$,
a dipolar BEC is well described by the macroscopic wave function
$\psi:=\psi(\bx,t)$ whose evolution is governed by the following
3D GPE  \cite{BaoBenCai,Baocai2013,BaoCaiWang,Bar,Lahaye,Xiong}
\begin{equation*} \label{eq:ngpe:sec8}
i\hbar \partial_t\psi=\left[-\frac{\hbar^2}{2m}\nabla^2+V(\bx)+Ng|\psi|^2+
NC_{dd}\left(V_{\rm dip}\ast |\psi|^2\right)\right]\psi, \quad \bx\in{\Bbb
R}^3, \ t>0,
\end{equation*}
where $C_{dd}=\mu_0\mu_{\rm dip}^2/3$ with $\mu_0$ the vacuum magnetic permeability and
$\mu_{\rm dip}$ the permanent magnetic dipole moment, $\psi$
satisfies the normalization condition (\ref{GPnorm}), and
the long-range and anisotropic DDI between two dipoles with the same
dipole moment or orientation ${\bf n}=(n_1,n_2,n_3)^T\in {\Bbb R}^3$ (which is
a given unit vector satisfying $|{\bf n}|=\sqrt{n_1^2+n_2^2+n_3^3}=1$) is given by
\begin{equation}
\label{eq:kel0:sec8} V_{\rm dip}(\bx)= \frac{3}{4\pi}\,\frac{1-3(\bx\cdot \bf
n)^2/|\bx|^2}{|\bx|^3}=\frac{3}{4\pi}\,\frac{1-3\cos^2(\theta)}{|\bx|^3}, \qquad \bx\in{\Bbb
R}^3,\end{equation}
where  $\theta$ is the angle
between the dipole axis ${\bf n}$ and the vector $\bx$. We remark here that
it is still an open problem to derive the above GPE from the $N$-body linear Schr\"{o}dinger
equation (\ref{eq:Nbdy}) with $V_{\rm int}$ in (\ref{energy-total}) is taken as $V_{\rm dip}$.

Again, under the harmonic potential (\ref{eq:hp}), similarly to
the nondimensionalization in section 2.2 and dimension reduction
in 2.3 from 3D to 2D when $\omega_z\gg \max\{\omega_x,\omega_y\}$ for a disk-shaped condensate
and to 1D when $\omega_z=\omega_y\gg \omega_x$  for a cigar-shaped condensate
\cite{BaoBenCai,Baocai2013,CaiRosen},
 by using the decomposition of contact and long-range
(or repulsive and attractive) parts of the DDI (\ref{eq:kel0:sec8})  \cite{BaoCaiWang,CaiRosen}
\begin{equation} \label{eq:decop1:sec8} U_{\rm dip}(\bx)=\frac{3}{4\pi
|\bx|^3}\left(1-\frac{3(\bx\cdot {\bf n})^2}{|\bx|^2}\right) = -
\delta ({\bf x})-3\partial_{{\bf n}{\bf n}}\left( \frac{1}{4\pi |\bx|}\right),\quad
\bx\in {\Bbb R}^3,\end{equation}
where the differential operators $\partial_{\bf n}={\bf n}\cdot\nabla$ and
$\partial_{{\bf n} {\bf n}}=\partial_{\bf n}\partial_{\bf n}$,
we can obtain the following dimensionless GPE with a DDI
in $d$-dimensions ($d=1,2,3$):
\begin{equation}\label{GGPE}
\begin{split}
&i\partial_t\psi({\bx}, t) = \left[-\frac{1}{2}\nabla^2 + V({\bf x}) +
\beta|\psi(\bx,t)|^2 +\eta \varphi(\bx,t)\right]\psi(\bx,t), \\
&\varphi(\bx,t)=L_{\bf n} u(\bx,t), \qquad u(\bx,t)=G*|\psi|^2,
\qquad\qquad \qquad \bx\in{\mathbb R}^d, \quad t\geq0,
\end{split}
\end{equation}
where
\begin{equation*}\label{Gpotential}
\beta = \left\{\begin{array}{l}
\frac{2\kappa +\lambda(1-3n_1^2)}{4\pi\varepsilon^2},\\
\frac{\kappa +\lambda(3n_3^2-1)}{\varepsilon\sqrt{2\pi}}, \\
\kappa - \lambda,
\end{array}\right.
\ \eta = -3\lambda\left\{\begin{array}{l}
\frac{3n_1^2-1}{8\vep\sqrt{2\pi}},\\
1/2, \\
1,
\end{array}\right. \ L_{\bf n} = \left\{\begin{array}{ll}
\partial_{xx}, &d=1,\\
 \partial_{\bf n_\bot n_\bot} - n_3^2\nabla^2, &d = 2,\\
\partial_{\bf nn}, &d = 3,
\end{array}\right.
\end{equation*}
with $\kappa= \frac{4\pi Na_s}{x_s}$, $\lambda
=\frac{mN\mu_0\mu_{\rm dip}^2}{3\hbar^2 x_s}$, $\vep=\frac{1}{\sqrt{\gamma_z}}$, ${\bf n}_\perp=(n_1,n_2)^T$, and
\begin{equation*}\label{poten987}
G(\bx)=\left\{\begin{array}{l}
\frac{1}{\vep\sqrt{2\pi}}\int_0^\infty\frac{e^{-s/2\vep^2}}{\sqrt{s^2+|\bx|^2}}ds\\
1/(2\pi|\bx|), \\
\frac{1}{(2\pi)^{3/2}}\int_{\mathbb R}
\frac{e^{-s^2/2}}{\sqrt{|\bx|^2+\varepsilon^2s^2}}ds, \\
1/(4\pi|\bx|), \\
\end{array}\right.
\Leftrightarrow \widehat{G}(\xi)=\left\{\begin{array}{ll}
\frac{\vep\sqrt{2}}{\sqrt{\pi}}\int_0^\infty
\frac{e^{-\vep^2 s/2}}{s+|\xi|^2}ds, &d=1\&\hbox{SAM},\\
1/|\xi|, &d = 2\&\hbox{SDM},\\
\frac{1}{2\pi^2}\int_{\mathbb R}\frac{e^{-\varepsilon^2 s^2/2}}{|\xi|^2+s^2}ds, &d = 2\&\hbox{SAM},\\
1/|\xi|^2, &d=3,\\
\end{array}\right.
\end{equation*}
where $\widehat{f}(\xi)$ denotes the Fourier transform of a function $f(\bx)$ for $\bx,\,
\xi\in {\mathbb R}^d$. In addition, in 3D, $u$ in (\ref{GGPE}) satisfies the
Poisson equation \cite{BaoBenCai,Baocai2013,CaiRosen}
\begin{equation}
-\nabla^2 u(\bx,t)=|\psi(\bx,t)|^2, \quad \bx\in{\Bbb R}^3,  \quad \hbox{satisfying}\ \
\lim_{|\bx|\to \infty} u(\bx,t)=0, \quad  t\ge0;
\end{equation}
and in 2D with SDM approximation, $u$ in (\ref{GGPE}) satisfies the square-root-Poisson equation \cite{BaoBenCai,Baocai2013,CaiRosen}
\begin{equation}
(-\nabla^2)^{1/2} u(\bx,t)=|\psi(\bx,t)|^2, \ \bx\in{\Bbb R}^2,  \ \hbox{satisfying}\ \
\lim_{|\bx|\to \infty} u(\bx,t)=0, \  t\ge0.
\end{equation}
The GPE (\ref{GGPE}) conserves the
normalization (\ref{GPnorm}) and energy per particle
\begin{equation*}\label{eq:enfg1:sec5}
E(\psi(\cdot,t)) =\int_{{\Bbb
R}^d} \left[\frac{1}{2} \left|\nabla \psi\right|^2+
V(\bx)|\psi|^2 +\frac{\beta}{2}\,
|\psi|^4+\frac{\eta}{2} \varphi|\psi|^2\right]d\bx\equiv E(\psi(\cdot,0)),\ t\ge0.
\end{equation*}

The ground state can be defined the same as
(\ref{eq:minp:sec2}) with the above energy functional.
For the existence and uniqueness as well as nonexistence of the ground state of
the dipolar BEC (\ref{GGPE}) and
efficient and accurate numerical methods  for simulation, such as BESP \cite{BaoCaiWang}
or BEFP with nonuniform FFT \cite{Jiang},
we refer to \cite{BaoBenCai,Baocai2013,BaoCaiWang} and references therein.
Similarly, for the well-posedness of the Cauchy problem of (\ref{GGPE})
with the initial data (\ref{initial}) and its dynamical properties as well
as efficient and accurate numerical methods, such as TSSP \cite{BaoCaiWang}
or TSFP with nonuniform FFT \cite{Jiang},
we refer to \cite{BaoBenCai,Baocai2013,BaoCaiWang} and references therein.

\subsection{For spin-orbit-coupled BEC}
At temperatures $T$ much smaller than the critical temperature
$T_c$,  a spin-orbit-coupled BEC with two components
can be well described by
the macroscopic wave function $\Psi:=\Psi(\bx,t)=(\psi_1(\bx,t),\psi_2(\bx,t))^T$
whose evolution is governed by the following 3D coupled Gross-Pitaevskii equations (CGPEs)
\cite{Bao,Baocai2013,BaoCai0,Jaksch,Lin2011,Pethick,PitaevskiiStringari,ZhangBaoLi} for
$\bx\in\Bbb{R}^3$ and $t>0$ as
\begin{equation*}\label{eq:cgpe109:sec9}
\begin{split}
&i\hbar\partial_t \psi_1=\left[-\frac{\hbar^2}{2m}\nabla^2
+V(\bx)+\frac{i\hbar\tilde{k}_0}{2m}\partial_x+\frac{\hbar \tilde{\delta}}{2}
+Ng_{11}|\psi_1|^2+Ng_{12}|\psi_2|^2\right]\psi_1+\frac{\hbar\tilde{\Omega}}{2}
\psi_2, \\
&i\hbar\partial _t \psi_2=\left[-\frac{\hbar^2}{2m}\nabla^2
+V(\bx)-\frac{i\hbar\tilde{k}_0}{2m}\partial_x-\frac{\hbar \tilde{\delta}}{2}
+Ng_{21}|\psi_1|^2+Ng_{22}|\psi_2|^2\right]
\psi_2+\frac{\hbar\tilde{\Omega}}{2}
\psi_1,
\end{split}
\end{equation*}
where $N$ is the total number of particles,
$\tilde{k}_0$ describes the spin-orbit-coupling strength,
$\tilde{\delta}$ is the detuning constant for  Raman transition,
$\tilde{\Omega}$ is the effective
Rabi frequency describing the strength to realize the internal atomic
Josephson junction (JJ) by a Raman transition, and
the interactions of particles are described by
$g_{jl}=\frac{4\pi \hbar^2 a_{jl}}{m}$ with $a_{jl}=a_{lj}$ ($j,l=1,2$) being
the $s$-wave scattering lengths between the $j$th and $l$th components.
The above CGPEs is normalized as
\begin{equation}\label{socbec}
\|\Psi\|^2:=\int_{{\Bbb R}^3}
\left[|\psi_1(\bx,t)|^2+|\psi_2(\bx,t)|^2\right]\,d\bx=1.
\end{equation}

Again, under the harmonic potential (\ref{eq:hp}), similarly to
the nondimensionalization in section 2.2 and dimension reduction
in 2.3 from 3D to 2D and 1D, we can obtain the following
dimensionless CGPEs under the normalization condition (\ref{socbec})
for spin-orbit-coupled BEC
in $d$-dimensions ($d=1,2,3$) for $\bx\in{\Bbb R}^d$ and $t>0$ as
\begin{equation}\label{eq:cgpe1569:sec9}
\begin{split}
&i\partial_t \psi_1=\left[-\frac{1}{2}\nabla^2
+V(\bx)+i k_0\partial_x+\frac{\delta}{2}
+\beta_{11}|\psi_1|^2+\beta_{12}|\psi_2|^2\right]\psi_1+\frac{\Omega}{2}
\psi_2, \\
&i\partial _t \psi_2=\left[-\frac{1}{2}\nabla^2
+V(\bx)-ik_0\partial_x-\frac{\delta}{2}
+\beta_{21}|\psi_1|^2+\beta_{22}|\psi_2|^2\right]
\psi_2+\frac{\Omega}{2}
\psi_1,
\end{split}
\end{equation}
where $k_0=\frac{\tilde{k}_0}{\omega_x}$, $\delta=\frac{\tilde{\delta}}{\omega_x}$,
$\Omega=\frac{\tilde{\Omega}}{\omega_x}$, and $\beta_{11}$, $\beta_{12}=\beta_{21}$,
$\beta_{22}$ are dimensionless interaction constants. This CGPEs conserves
the normalization (or total mass)
\begin{equation}
\label{eq:mass1:sec9}
N(\Psi(\cdot,t)):=\|\Psi(\cdot,t)\|^2=\int_{{\Bbb
R}^d} \sum_{j=1}^2|\psi_j(\bx,t)|^2\,d\bx\equiv N(\Psi(\cdot,0))
=1, \quad t\ge0, \end{equation}
and the energy per particle
\begin{align}
&E(\Psi(\cdot,t))=\int_{{\Bbb R}^d}\biggl\{\sum_{j=1}^2\biggl[\frac
12|\nabla\psi_j|^2+|\psi_j|^2\left(V(\bx)+\frac 12\sum_{l=1}^2\beta_{jl}|\psi_l|^2\right)\biggr]+
\frac{\delta}{2}\left(|\psi_1|^2-|\psi_2|^2\right)\nonumber\\
&\qquad +ik_0\left(\overline{\psi}_1\partial_x\psi_1
-\overline{\psi}_2\partial_x\psi_2\right)
+\Omega \, \text{Re}(\psi_1\overline{\psi}_2)\biggl\}d\bx\equiv E(\Psi(\cdot,0)),\quad t\ge0.
\label{eq:energy:sec9}
\end{align}
In addition, when $\Omega=0$, then it also conserves the mass of each component
 \begin{equation}
\label{eq:masggs1:sec9}
N(\psi_j(\cdot,t)):=\int_{{\Bbb
R}^d} |\psi_j(\bx,t)|^2\,d\bx\equiv N(\psi_j(\cdot,0)), \quad t\ge0, \quad
j=1,2.\end{equation}

The ground state can be defined as:  Find $ \Phi_g\in S$ such that
\begin{equation}
\label{eq:minp:sec687}
E_g:=E(\Phi_g) = \min_{\Phi\in S} E(\Phi),
\end{equation}
where $S=\{\Phi=(\phi_1,\phi_2)^T \ |\ \|\Phi\|=1, \ E(\Phi)<\infty\}$.
Of course, when $\Omega=0$, for any fixed $0\le \alpha\le 1$, an $\alpha$-dependent
ground state can be defined as:
 Find $\Phi_g^\alpha\in S_\alpha$ such that
\begin{equation}
\label{eq:minp:sec8787}
E_g^\alpha:=E(\Phi_g^\alpha) = \min_{\Phi\in S_\alpha} E(\Phi),
\end{equation}
where $S_\alpha=\{\Phi=(\phi_1,\phi_2)^T \ |\ \|\phi_1\|^2=\alpha, \|\phi_1\|^2=1-\alpha, \ E(\Phi)<\infty\}$.
It is easy to see that
\begin{equation} \label{eq:minp:se8797}
E_g=E(\Phi_g)=\min_{0\le \alpha\le 1} E_g^\alpha
=\min_{0\le \alpha\le 1} E(\Phi_g^\alpha) =\min_{0\le \alpha\le 1}\min_{\Phi\in S_\alpha} E(\Phi).
\end{equation}
For the existence and uniqueness as well as nonexistence of the ground states of
the spin-orbit-coupled BEC (\ref{eq:cgpe109:sec9}) based on the definition (\ref{eq:minp:sec687})
for any $\Omega\in {\Bbb R}$
and the definition (\ref{eq:minp:sec8787}) for $\Omega=0$, and
efficient and accurate numerical methods  for simulation, such as BEFD or BESP \cite{Bao,Baocai2013,BaoCai0},
we refer to \cite{Bao,BaoCai0,Baocai2013,Pethick,PitaevskiiStringari,ZhangBaoLi} and references therein.
Similarly, for the well-posedness of the Cauchy problem of (\ref{eq:cgpe109:sec9})
with the initial data $\Psi(\bx,0)=\Psi_0(\bx)$
and its dynamical properties as well
as efficient and accurate numerical methods, such as TSSP \cite{Bao,Baocai2013},
we refer to \cite{Bao,Baocai2013,BaoCai0,Pethick,PitaevskiiStringari,ZhangBaoLi} and references therein.
Finally, by setting $\psi_1(\bx,t)=\varphi_1(\bx,t)
 e^{i(\omega t+k_0x)}$ and $\psi_2(\bx,t)=\varphi_2(\bx,t)
 e^{i(\omega t-k_0x)}$ with $\omega=\frac{\delta-k_0^2}{2}$
in the CGPEs (\ref{eq:cgpe1569:sec9}), we obtain for $\bx\in{\Bbb R}^d$ and $t>0$
\begin{equation}\label{eq:cgpe199:sec9}
\begin{split}
&i\partial_t \varphi_1=\left[-\frac{1}{2}\nabla^2
+V(\bx)+\delta
+\beta_{11}|\varphi_1|^2+\beta_{12}|\varphi_2|^2\right]\varphi_1+\frac{\Omega}{2}
e^{-i2k_0x}\varphi_2, \\
&i\partial _t \varphi_2=\left[-\frac{1}{2}\nabla^2
+V(\bx)+\beta_{21}|\varphi_1|^2+\beta_{22}|\varphi_2|^2\right]
\varphi_2+\frac{\Omega}{2}e^{i2k_0x}
\varphi_1.
\end{split}
\end{equation}
This CGPEs conserves the normalization (\ref{eq:mass1:sec9}) for any $\Omega\in {\Bbb R}$
and (\ref{eq:masggs1:sec9})
when $\Omega=0$ with $\psi_j$ replaced by $\varphi_j$ for $j=1,2$. It is very useful in designing
the most efficient and accurate numerical methods
for computing ground states and dynamics, such as BESP and TSSP \cite{Bao,Baocai2013,BaoCai0}),
especially for the box potential.

\section{Conclusions and future perspectives}

Due to its massive relations and applications in many different areas,
such as atomic, molecular and optical physics, quantum optics,
condense matter physics and low temperature physics,
the research on theoretical, experimental and computational studies
of BEC has been started almost century ago and has grown explosively (or exponentially)
since 1995. Up to now, rich and extensive research results
have been obtained in experimental and theoretical understanding of
ground states and dynamics of BEC. The research in this area
is still very active and highly demanded due to the latest
experimental and/or technological advances in BEC,
such as spinor BEC \cite{BaoLim,BaoWang2,Ho1,Ueda}, BEC with damping terms \cite{BaoJakschP46}
or impurities \cite{Jon} or random potentials \cite{Min},
 degenerate Fermi gas \cite{GiorginiPS}, Rydberg gas \cite{DietJ},
spin-orbit-coupled BEC \cite{Lin2011}, BEC at finite temperature \cite{Zaremba}, etc.
These achievements have brought great challenges to AMO community,
condensed matter community, and computational and applied mathematics community
for modeling, simulating and understanding various interesting phenomenons related to BEC.
It becomes more and more interdisciplinary involving
theoretical, computational and experimental physicists and
computational and applied mathematicians as well as pure mathematicians.


\begin{thebibliography}{99}


\bibitem{Abo}  Abo-Shaeer, J. R., Raman, C., Vogels J. M., Ketterle, W.,
Observation of vortex lattices in Bose-Einstein condensates,
\emph{Science} \textbf{292} (2001),  476--479.


\bibitem{Aftalion}
Aftalion, A.,
\emph{Vortices in Bose-Einstein Condensates}.
Progress in Nonlinear Differential Equations and their Applications, 67, Birkh\"auser, Boston, 2006.





\bibitem{Anderson}
 Anderson, M. H.,  Ensher, J. R.,  Matthewa, M. R.,  Wieman C. E.,
Cornell, E. A., Observation of Bose-Einstein condensation in a dilute
atomic vapor, \emph{Science} \textbf{269} (1995),  198--201.


\bibitem{Ant} Antoine, X.,  Bao, W., Besse, C.,
Computational methods for the dynamics of the nonlinear Schr\"{o}dinger/Gross-Pitaevskii equations,
\emph{Comput. Phys. Commun.} \textbf{184} (2013),  2621-2633.

\bibitem{Bao}  Bao, W.,
Ground states and dynamics of multicomponent Bose-Einstein condensates,
\emph{Multiscale Model. Simul.} \textbf{2} (2004), 210--236.




\bibitem{BaoBenCai} Bao, W., Ben Abdallah N.,  Cai, Y.,
Gross-Pitaevskii-Poisson equations for dipolar Bose-Einstein condensate with anisotropic confinement,
\emph{SIAM J. Math. Anal.} \textbf{44} (2012),  1713--1741.

\bibitem{Baocai2013} Bao, W., Cai, Y.,
Mathematical theory and numerical methods for Bose-Einstein condensation,
\emph{Kinet. Relat. Mod.} \textbf{6} (2013),  1-135.

\bibitem{BaoCai0} Bao W.,  Cai, Y.,
Ground states of two-component Bose-Einstein condensates with an internal atomic Josephson junction,
\emph{East Asia J.  Appl. Math.} \textbf{1} (2010), 49--81.


\bibitem{BaoCai2} Bao W., Cai, Y.,
Optimal error estimates of finite difference methods for the Gross-Pitaevskii equation
with angular momentum rotation,
\emph{Math. Comp.} \textbf{82} (2013), 99-129.



\bibitem{BaoCaiWang} Bao, W., Cai, Y., Wang, H.,
Efficient numerical methods for computing ground states and dynamics of
dipolar Bose-Einstein condensates,
\emph{J. Comput. Phys.} \textbf{229} (2010),  7874--7892.


\bibitem{BaoChernLim} Bao, W., Chern I-L., Lim, F. Y.,
Efficient and spectrally accurate numerical methods for
computing ground and first excited states in Bose-Einstein condensates,
\emph{J. Comput. Phys.} \textbf{219} (2006),  836--854.


\bibitem{BaoDu} Bao, W.,  Du, Q.,
Computing the ground state solution of Bose-Einstein condensates by a normalized gradient flow,
\emph{SIAM J. Sci. Comput.} \textbf{25} (2004), 1674--1697.


\bibitem{BaoDuZhang}  Bao, W.,  Du, Q., Zhang, Y.,
Dynamics of rotating Bose-Einstein condensates and its efficient and accurate numerical computation,
\emph{SIAM J. Appl. Math.} \textbf{66} (2006),  758--786.



\bibitem{BaoJakschP} Bao, W.,  Jaksch, D.,  Markowich, P. A.,
Numerical solution of the Gross-Pitaevskii equation for Bose-Einstein condensation,
\emph{J. Comput. Phys.}  \textbf{187} (2003), 318--342.

\bibitem{BaoJakschP46} Bao, W., Jaksch, D.,  Markowich, P. A.,
Three dimensional simulation of jet formation in collapsing condensates,
\emph{J. Phys. B: At. Mol. Opt. Phys.}  \textbf{37} (2004), 329--343.

\bibitem{BLM} Bao, W., Le Treust, L., M\'{e}hats, F.,
Dimension reduction for the anistropic Bose-Einstein condensates in
the strong interaction regime, preprint.


\bibitem{BaoLiShen}   Bao, W.,  Li, H. L.,  Shen, J.,
A generalized Laguerre-Fourier-Hermite pseudospectral method for
computing the dynamics of rotating Bose-Einstein condensates,
\emph{SIAM J. Sci. Comput.} \textbf{31} (2009),  3685--3711.

\bibitem{BaoLim} Bao, W.,  Lim, F. Y.,
Computing ground states of spin-1 Bose-Einstein
condensates by the normalized gradient flow,
\emph{SIAM J. Sci. Comput.} \textbf{30} (2008),  1925--1948.

\bibitem{Bao2013} Bao, W., Marahrens, D., Tang, Q., Zhang, Y.,
A simple and efficient numerical method for computing the dynamics of
rotating Bose-Einstein condensates via a rotating Lagrangian coordinate,
\emph{SIAM J. Sci. Comput.} \textbf{35} (2013), A2671-A2695.



\bibitem{BaoShen}  Bao, W.,  Shen, J.,
A fourth-order time-splitting Laguerre-Hermite pseudospectral method for Bose-Einstein condensates,
\emph{SIAM J. Sci. Comput.} \textbf{26} (2005), 2020--2028.




\bibitem{BaoWang}   Bao, W., Wang, H.,
An efficient and spectrally accurate numerical method for
computing dynamics of rotating Bose-Einstein condensates,
\emph{J. Comput. Phys.} \textbf{217} (2006),  612--626.

\bibitem{BaoWang2} Bao, W.,  Wang, H.,
A mass and magnetization conservative and energy-diminishing numerical
method for computing ground state of spin-1 Bose-Einstein condensates,
\emph{SIAM J. Numer. Anal.} \textbf{45} (2007), 2177--2200.


\bibitem{BaoWangP} Bao, W., Wang, H.,  Markowich, P. A.,
Ground, symmetric and central vortex states in rotating Bose-Einstein condensates,
\emph{Commun. Math. Sci.} \textbf{3} (2005),  57--88.


\bibitem{Bar}
Baranov, M. A.,
Theoretical progress in many-body physics with ultracold dipolar gases,
\emph{Phys. Rep.} \textbf{464} (2008), 71-111.

\bibitem{bamsw}
Ben Abdallah, N.,  M\'ehats, F., Schmeiser, C., Weish\"{a}upl, R. M.,
The nonlinear Schr\"{o}dinger equation with a strongly
anisotropic harmonic potential,
\emph{SIAM J. Math. Anal.} \textbf{37} (2005), 189--199.



\bibitem{BlochDZ} Bloch, I., Dalibard, J., Zwerger, W.,
Many-body physics with ultracold gases,
\emph{Rev. Mod. Phys.} \textbf{80} (2008),  885-964.




\bibitem{Bose}  Bose, S. N.,
Plancks gesetz und lichtquantenhypothese,
\emph{Zeitschrift fr Physik} \textbf{3} (1924), 178--181.


\bibitem{Bradley}  Bradley, C. C.,  Sackett, C. A.,  Tollett, J. J.,   Hulet, R. G.,
Evidence of Bose-Einstein condensation in an atomic
 gas with attractive interaction,
\emph{Phys. Rev. Lett.} \textbf{75} (1995), 1687--1690.



\bibitem{Bruderer} Bruderer, M., Bao, W.,  Jaksch, D.,
Self-trapping of impurities in Bose-Einstein condensates: Strong
attractive and repulsive coupling,
\emph{EPL} \textbf{82} (2008), 30004.





\bibitem{CaiRosen}  Cai, Y., Rosenkranz, M.,  Lei, Z.,  Bao, W.,
Mean-field regime of trapped dipolar Bose-Einstein condensates in
one and two dimensions,
\emph{Phys. Rev. A} \textbf{82} (2010),  043623.


\bibitem{Caradoc}  Caradoc-Davis, B. M.,  Ballagh, R. J., Burnett, K.,
Coherent dynamics of vortex formation in trapped Bose-Einstein condensates,
\emph{Phys. Rev. Lett.} \textbf{83} (1999),  895--898.


\bibitem{Carles01}
Carles, R.,
\emph{Semi-Classical Analysis for Nonlinear Schr\"{o}dinger Equations},
World Scentific,  2008.




\bibitem{Carles2}  Carles, R.,  Markowich, P. A., Sparber, C.,
On the Gross--Pitaevskii equation for trapped dipolar quantum gases,
\emph{Nonlinearity} \textbf{21} (2008),  2569--2590.





\bibitem{Cerim}  Cerimele, M. M.,  Pistella, F.,  Succi, S.,
Particle-inspired scheme for the Gross-Pitaevski equation: An application to Bose-Einstein
condensation,
\emph{Comput. Phys. Comm.} \textbf{129} (2000),  82--90.

\bibitem{ChenT}
Chen T., Pavlovi\'c N.,
Derivation of the cubic NLS and Gross-Pitaevskii hierarchy from many body dynamics in $d=3$
based on space time norms,
\emph{ Ann. H. Poincare}, \textbf{15} (2014),  543--588.

\bibitem{ChenX}
Chen X.,
On the rigorous derivation of the 3D cubic nonlinear Schr\"odinger equation with a quadratic trap,
\emph{Arch. Rational Mech. Anal.}, \textbf{210} (2013),  365--408.

\bibitem{Cornell}   Cornell, E. A.,   Wieman, C. E.,
Nobel Lecture: Bose-Einstein condensation in a dilute gas,
the first 70 years and some recent experiments,
\emph{Rev. Mod. Phys.} \textbf{74} (2002), 875--893.






\bibitem{Dalfovo} Dalfovo, F.,  Giorgini, S.,  Pitaevskii, L. P.,  Stringari, S.,
Theory of Bose-Einstein condensation in trapped gases,
\emph{Rev. Mod. Phys.} \textbf{71} (1999), 463--512.


\bibitem{Davis}   Davis, K. B.,  Mewes, M. O., Andrews,  M. R.,  van Druten, N. J.,
  Durfee, D. S.,  Kurn, D. M.,  Ketterle, W.,
Bose-Einstein condensation in a gas of sodium atoms,
\emph{Phys. Rev. Lett.} \textbf{75} (1995),  3969--3973.


\bibitem{Einstein1}
Einstein, A.,
Quantentheorie des einatomigen idealen gases,
\emph{Sitzungsberichte der Preussischen Akademie der Wissenschaften} \textbf{22} (1924),  261--267.


\bibitem{Einstein2} Einstein, A.,
Quantentheorie des einatomigen idealen gases, zweite abhandlung,
\emph{Sitzungs-berichte der Preussischen Akademie der Wissenschaften}
\textbf{1} (1925),  3--14.

\bibitem{Erdos}   Erd\H{o}s, L., Schlein, B.,   Yau, H. T.,
Derivation of the Gross-Pitaevskii
equation for the dynamics of Bose-Einstein condensate,
\emph{Ann. Math.} \textbf{172} (2010),  291--370.

\bibitem{Fetter}   Fetter, A. L.,
Rotating trapped Bose-Einstein condensates,
\emph{Rev. Mod. Phys.} \textbf{81} (2009),  647--691.


\bibitem{GMMP}
Gerard, P., Markowich, P. A.,  Mauser, N. J.,  Poupaud, F.,
Homogenization limits and Wigner transforms,
\emph{Comm. Pure Appl. Math.} \textbf{50} (1997),  321--377.



\bibitem{GiorginiPS} Giorgini, S.,  Pitaevskii, L. P.,  Stringari, S.,
Theory of ultracold atomic Fermi gases,
\emph{Rev. Mod. Phys.} \textbf{80} (2008),  1215--1274.



\bibitem{Gross} Gross, E. P.,
Structure of a quantized vortex in boson systems,
\emph{Nuovo. Cimento.} \textbf{20} (1961),  454--457.


\bibitem{Ho1}  Ho, T. L.,
Spinor Bose condensates in optical traps,
\emph{Phys. Rev. Lett.} \textbf{81} (1998),  742--745.

\bibitem{Jaksch}
Jaksch, D., Gardiner, S. A., Schulze, K., Cirac J. I.,  Zoller, P.,
Uniting Bose-Einstein condensates in optical resonators,
\emph{Phys. Rev. Lett.} \textbf{86} (2001),  4733--4736.



\bibitem{Jiang} Jiang, S., Greengard, L., Bao, W.,
Fast and accurate evaluation of nonlocal Coulomb and dipole-dipole
interactions via the nonuniform FFT,
\emph{arXiv: 1311.4120} (2013).

\bibitem{Jon}  Johnson, T. J., Bruderer, M., Cai, Y.,   Clark, S. R.,  Bao, W.,  Jaksch, D.,
Breathing oscillations of a trapped impurity in a Bose gas,
\emph{EPL} \textbf{98} (2012), article 26001.



\bibitem{Ueda}  Kawaguchi Y.,  Ueda, M.,
Spinor Bose-Einstein condensates,
\emph{Phys. Rep.}, \textbf{520} (2012), 253-381.

\bibitem{Ketterle}  Ketterle, W.,
Nobel lecture: When atoms behave as waves: Bose-Einstein
condensation and the atom laser,
\emph{Rev. Mod. Phys.} \textbf{74} (2002), 1131--1151.

\bibitem{DietJ}
Kiffner, M., Li, W., Jaksch, D.,
Three-body bound states in dipole-dipole interacting Rydberg atoms,
\emph{Phys. Rev. Lett.} \textbf{111} (2013), 233003.

\bibitem{Klainer}
 Klainerman S.,  Machedon M.,
On the uniqueness of solutions to the Gross-Pitaevskii hierarchy,
\emph{Commun. Math. Phys.} \textbf{279} (2008), 169--185 .


\bibitem{Lahaye}
Lahaye, T., Menotti, C., Santos, L., Lewenstein, M., Pfau, T.,
The physics of dipolar bosonic quantum gases,
\emph{Rep. Prog. Phys.} \textbf{72} (2009) 126401.

\bibitem{Leggett} Leggett, A. J.,
Bose-Einstein condensation in the alkali gases: Some fundamental concepts,
\emph{Rev. Mod. Phys.} \textbf{73} (2001), 307--356.


\bibitem{LiebSeiringer1}   Lieb, E. H.,   Seiringer, R.,
Derivation of the
Gross-Pitaevskii equation for rotating Bose gases,
\emph{Comm. Math. Phys.} \textbf{264} (2006),  505--537.



\bibitem{LiebSeiringer}   Lieb, E. H., Seiringer, R.,  Solovej, J. P.,  Yngvason, J.,
\emph{The Mathematics of the Bose Gas and its Condensation}.
Oberwolfach Seminars 34, Birkh\"auser Verlag, Basel, 2005.

\bibitem{LiebSeiringerPra2000}   Lieb, E. H., Seiringer, R.,   Yngvason, J.,
Bosons in a trap: A rigorous derivation of the Gross-Pitaevskii energy functional,
\emph{Phys. Rev. A} \textbf{61} (2000), 043602.

\bibitem{Lin2011}
Lin, Y. J., Jim\'enez-Garcia, K.,  Spielman, I. B.,
Spin-orbit-coupled Bose-Einstein condensates,
\emph{Nature}  \textbf{471} (2011), 83--86.

\bibitem{London}  London, F.,
The $\lambda$-phenomenon of liquid helium and the Bose-Einstein degeneracy,
\emph{Nature} \textbf{141} (1938),  643--644.



\bibitem{Matthews}  Matthews, M. R.,  Anderson, B. P.,  Haljan, P. C.,
 Hall, D. S.,  Wieman, C. E.,  Cornell, E. A.,
Vortices in a Bose-Einstein condensate,
\emph{Phys. Rev. Lett.} \textbf{83} (1999), 2498--2501.


\bibitem{Min} Min, B., Li, T., Rosenkranz, M., Bao, W.,
Subdiffusive spreading of a Bose-Einstein condensate in random potentials,
\emph{Phys. Rev. A} \textbf{86} (2012), article 053612.


\bibitem{Ming}
Minguzzi, A., Succi, S., Toschi, F., Tosi, M. P., Vignolo, P.,
Numerical methods for atomic quantum gases with applications
to Bose-Einstein condensates and to ultracold fermions,
\emph{Phys. Rep.} \textbf{395} (2004), 223-355.

\bibitem{Pethick}   Pethick, C. J., Smith, H.,
\emph{Bose-Einstein Condensation in Dilute Gases}.
 Cambridge University Press, 2002.

\bibitem{Pitaevskii}   Pitaevskii, L. P.,
Vortex lines in an imperfect Bose gas,
\emph{Soviet Phys. JETP} \textbf{13} (1961),  451--454.

\bibitem{PitaevskiiStringari}   Pitaevskii, L. P.,  Stringari, S.,
\emph{Bose-Einstein Condensation}.
Clarendon Press, Oxford, 2003.

\bibitem{Rosen} Rosenkranz, M., Jaksch, D.,  Lim, F. Y., Bao, W.,
Self-trapping of Bose-Einstein condensates expanding in shallow optical lattices,
\emph{Phys. Rev. A}  \textbf{77} (2008) article 063607.



\bibitem{Seiringer}   Seiringer, R.,
Gross-Pitaevskii theory of the rotating Bose gas,
\emph{Comm. Math. Phys.} \textbf{229} (2002),  491--509.


\bibitem{Sulem}
Sulem, C.,  Sulem, P. L.,
\emph{The Nonlinear Schr\"{o}dinger Equation,
Self-focusing and Wave Collapse},
 Springer-Verlag, New York, 1999.



\bibitem{Xiong} Xiong, B.,  Gong, J.,  Pu, H.,  Bao, W.,  Li, B.,
Symmetry breaking and
self-trapping of a dipolar Bose-Einstein condensate in a double-well
potential,
\emph{Phys. Rev. A} \textbf{79} (2009), 013626.




\bibitem{Zaremba}  Zaremba, E.,  Nikuni, T.,   Griffin, A.,
Dynamics of trapped Bose gases at finite temperature,
\emph{J. Low Temp. Phys.} \textbf{116} (1999),  277.


\bibitem{ZhangBaoLi}  Zhang, Y.,  Bao, W.,  Li, H. L.,
Dynamics of rotating two-component Bose-Einstein
condensates and its efficient computation,
\emph{Phys. D} \textbf{234} (2007),  49--69.







\end{thebibliography}
\end{document}